\documentclass[english]{achemso}
\usepackage[T1]{fontenc}
\usepackage[utf8]{inputenc}
\usepackage{amsmath}
\usepackage{amssymb}
\usepackage{stmaryrd}
\usepackage{graphicx}
\usepackage{pdfpages}

\makeatletter

\title{Fewest-Switches Surface Hopping with Long Short-Term Memory Networks}

\author{Diandong Tang}

\author{Luyang Jia}

\affiliation{Key Laboratory of Theoretical and Computational Photochemistry of
Ministry of Education, College of Chemistry, Beijing Normal University,
Beijing 100875, China}

\author{Lin Shen}

\email{lshen@bnu.edu.cn}

\affiliation{Key Laboratory of Theoretical and Computational Photochemistry of
Ministry of Education, College of Chemistry, Beijing Normal University,
Beijing 100875, China}

\alsoaffiliation{Yantai-Jingshi Institute of Material Genome Engineering, Yantai 265505
Shandong, China}

\author{Wei-Hai Fang}

\affiliation{Key Laboratory of Theoretical and Computational Photochemistry of
Ministry of Education, College of Chemistry, Beijing Normal University,
Beijing 100875, China}

\alsoaffiliation{Yantai-Jingshi Institute of Material Genome Engineering, Yantai 265505
Shandong, China}

\makeatother

\usepackage{babel}
\begin{document}
\pagebreak{}
\begin{abstract}
The mixed quantum-classical dynamical simulation is essential to study
nonadiabatic phenomena in photophysics and photochemistry. In recent
years, many machine learning models have been developed to accelerate
the time evolution of the nuclear subsystem. Herein, we implement
long short-term memory (LSTM) networks as a propagator to accelerate
the time evolution of the electronic subsystem during the fewest-switches
surface hopping (FSSH) simulations. A small number of reference trajectories
are generated using the original FSSH method, and then the LSTM networks
can be built, accompanied by careful examination of typical LSTM-FSSH
trajectories that employ the same initial condition and random numbers
as the corresponding reference. The constructed network is applied
to FSSH to further produce a trajectory ensemble to reveal the mechanism
of nonadiabatic processes. Taking Tully’s three models as test systems,
the collective results can be reproduced qualitatively. This work
demonstrates that LSTM is applicable to the most popular surface hopping
simulations.
\end{abstract}
\begin{center}
\includegraphics[width=2in,height=2in]{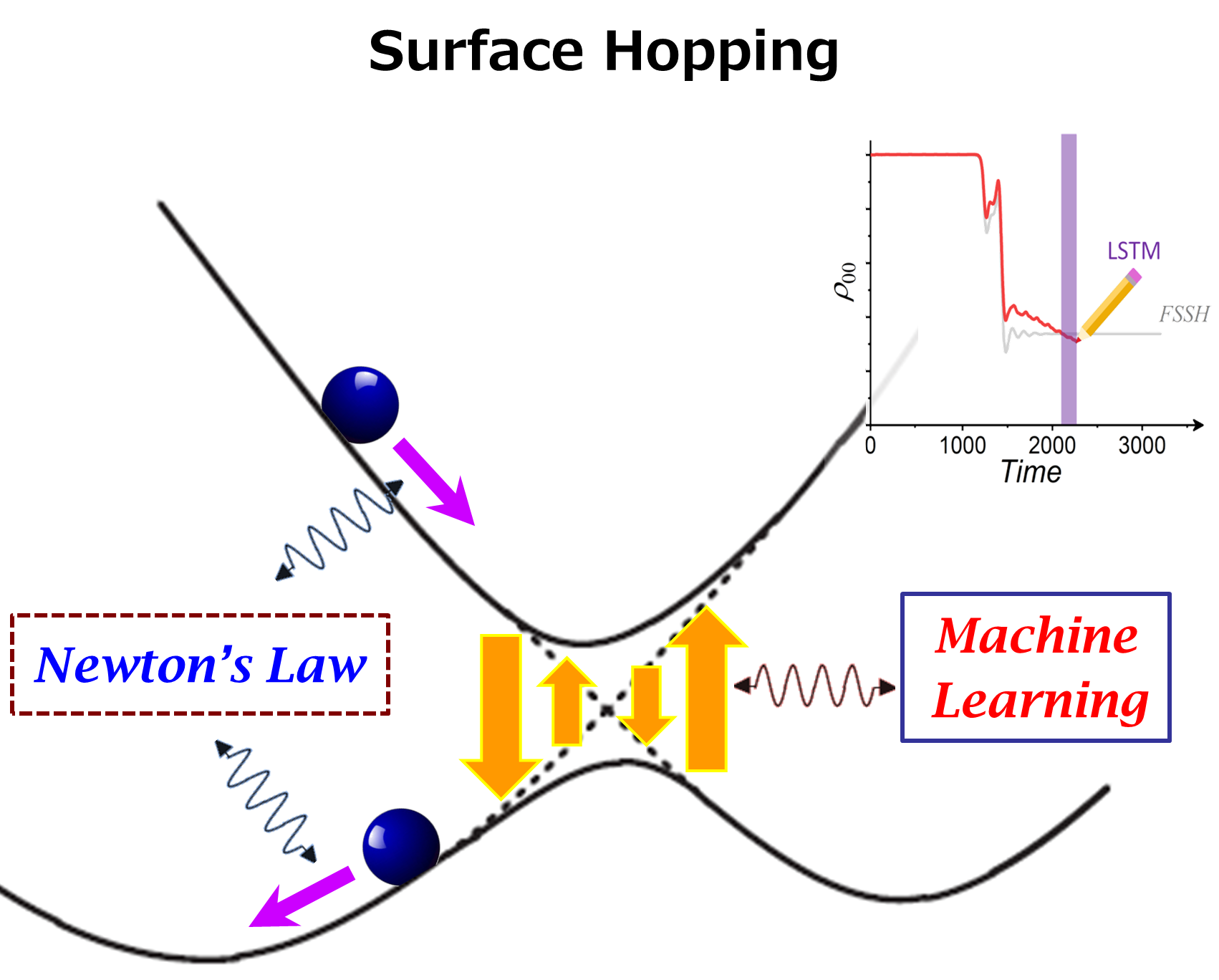}
\par\end{center}

\begin{center}
TOC graphic
\par\end{center}

\pagebreak{}

~~~~~Dynamical simulations governed by the laws of quantum mechanics
are indispensable for providing intuitive views of photophysical and
photochemical processes.\cite{Nelson_2020,Prezhdo_2021,Zobel_2021au}
Although a fully quantum treatment on both nuclear and electronic
motion should be employed in principle, such a simulation is too computationally
expensive for medium-sized or larger polyatomic molecules. The requirement
of prior knowledge about potential energy surface (PES) further hampers
its application on realistic molecular systems with high dimensionality.
As an alternative, the mixed quantum-classical molecular dynamics
(MQC-MD) approach provides a powerful tool to study the mechanism
of photochemical reactions, including the excited-state lifetimes
and the quantum yields of photoproducts on different reaction channels.
During MQC-MD simulations, the nuclei are represented as classical
particles but incorporate quantum feedback from the electronic motion.
Many theoretical models and algorithms for MQC-MD have been developed
over the past decades.\cite{Tully_1998,Stock_2005,Curchod_2013,Crespo_Otero_2018,Gao_2018,Mai_2020}
The most popular method is the fewest-switches surface hopping (FSSH)\cite{Tully_1990,Hammes_Schiffer_1994},
in which the nuclei travel classically on a single PES in an adiabatic
electronic state with a probability of switching to any other state.
The hopping probability is extracted from the time evolution of electronic
degrees of freedom. Despite its simplicity for implementation and
great success in many applications\cite{Shen_2020,Zobel_2021,Xie_2022},
the computational cost on FSSH is still much more expensive than ground-state
MD, especially with the growing demand of simulations on photodynamics
in long timescale.\cite{Mukherjee_2022} At each time step, electronic
structure calculations on ground and excited electronic states, including
potential energies, gradients as well as nonadiabatic coupling vectors
(NACVs) between different states, are usually required at a high level
of theory. How to accelerate dynamical simulations remains an important
topic if we are willing to extend FSSH to a broader prospect.

In recent years, we have witnessed the prosperity of machine learning
(ML) assisted chemical researches.\cite{Raccuglia_2016,Coley_2019,Zhou_2020,Zhang_2020,Dral_2020}
Several ML techniques such as kernel ridge regression and artificial
neural network have been proved as promising tools to provide a relation
between molecular structure and ground-state potential energy, so-called
ML-based PES or ML-based force field.\cite{Smith_2017,Chmiela_2018,Schutt_2018,Shen_2018,Zhang_2018,Unke_2019,Dral_2019,Behler_2021}
Because of its great potential to achieve high computational accuracy
and high efficiency on MD simulations simultaneously, the extension
to MQC-MD is attractive and seems to be straightforward. In principle,
each of adiabatic PESs can be individually fitted as the same as the
proposed ML-based force fields. However, the nonadiabatic coupling
vectors between adiabatic states have become a challenge for a fully
ML-based MQC-MD simulations. One solution is based on the Landau-Zener
formalism such as Zhu-Nakamura dynamics\cite{Zhu_1994,Zhu_1995},
in which the probability of surface hopping can be obtained only from
adiabatic energies and gradients, without the need to compute NACVs.
It has been successfully applied to ML-based nonadiabatic molecular
dynamics approaches recently\cite{Hu_2018,Chen_2018}, but the limitation
of Landau-Zener formalism as well as its influence on final simulation
results is not easy to realize or control. Another way is to improve
ML models to fit the reference values of NACVs.\cite{Dral_2018,Westermayr_2019,Westermayr_2020,Westermayr_2020cr}
Despite its success on some systems, it is worthy noting that machine
learning on nonadiabatic coupling is a nontrivial issue for high-dimensional
molecules because NACV keeps almost negligible in most regions of
PES but becomes very sensitive to small changes of nuclear positions
in the vicinity of conical intersections. The construction on a high-quality
ML database is also extremely time-consuming in practice on account
of multi-configuration electronic structure calculations for accurate
reference values.

In the above approaches, machine learning models are developed and
applied to predict potential energies, gradients and nonadiabatic
coupling vectors instead of electronic structure calculations, while
the time evolution of nuclei and electrons is still governed by the
Newton’s second law and the time-dependent Schrödinger equation, respectively.
A time series of nuclear positions, velocities and elements of electronic
density matrix is obtained and updated during simulations. It suggests
employing machine learning to propagate nuclear and/or electronic
subsystems based on historical knowledge extracted from the preceding
trajectory data. On one hand, ML techniques such as recurrent neural
network (RNN) and long short-term memory (LSTM) network have been
introduced to classical MD for realistic material and protein systems
to solve Newton’s equations.\cite{Wang_2020,Vlachas_2021,Kadupitiya_2022,Winkler_2022}
On the other hand, several successful ML applications on quantum dynamics
have been also proposed in recent years.\cite{Bandyopadhyay_2018,Herrera_Rodr_guez_2021,Reh_2021}
For example, Lan and coworkers simulated electronic evolution of multi-configuration
time-dependent hartree method with LSTM;\cite{Lin_2021} Ullah and
Dral studied quantum dissipative dynamics of spin-boson model and
Fenna-Matthews-Olson complex using kernel ridge regression and convolutional
neural network, respectively;\cite{Ullah_2021,Ullah_2022} Hammes-Schiffer
and coworkers employed artificial neural networks to solve the time-dependent
Schrödinger equation and propagate the wavepacket relevant to proton
transfer systems.\cite{Secor_2021} There are also reports of deep
learning in prediction of Landau-Zener transitions\cite{Yang_2020,Gao_2021}
as well as linearized semiclassical and symmetrical quasiclassical
mapping dynamics\cite{Wu_2021}. This inspired of us that machine
learning can be implemented as propagator for electronic degrees of
freedom in mixed quantum-classical molecular dynamics, possibly avoiding
expensive computation or difficult prediction on NACVs.

\begin{figure}
\noindent \centering{}\includegraphics[width=0.8\textwidth]{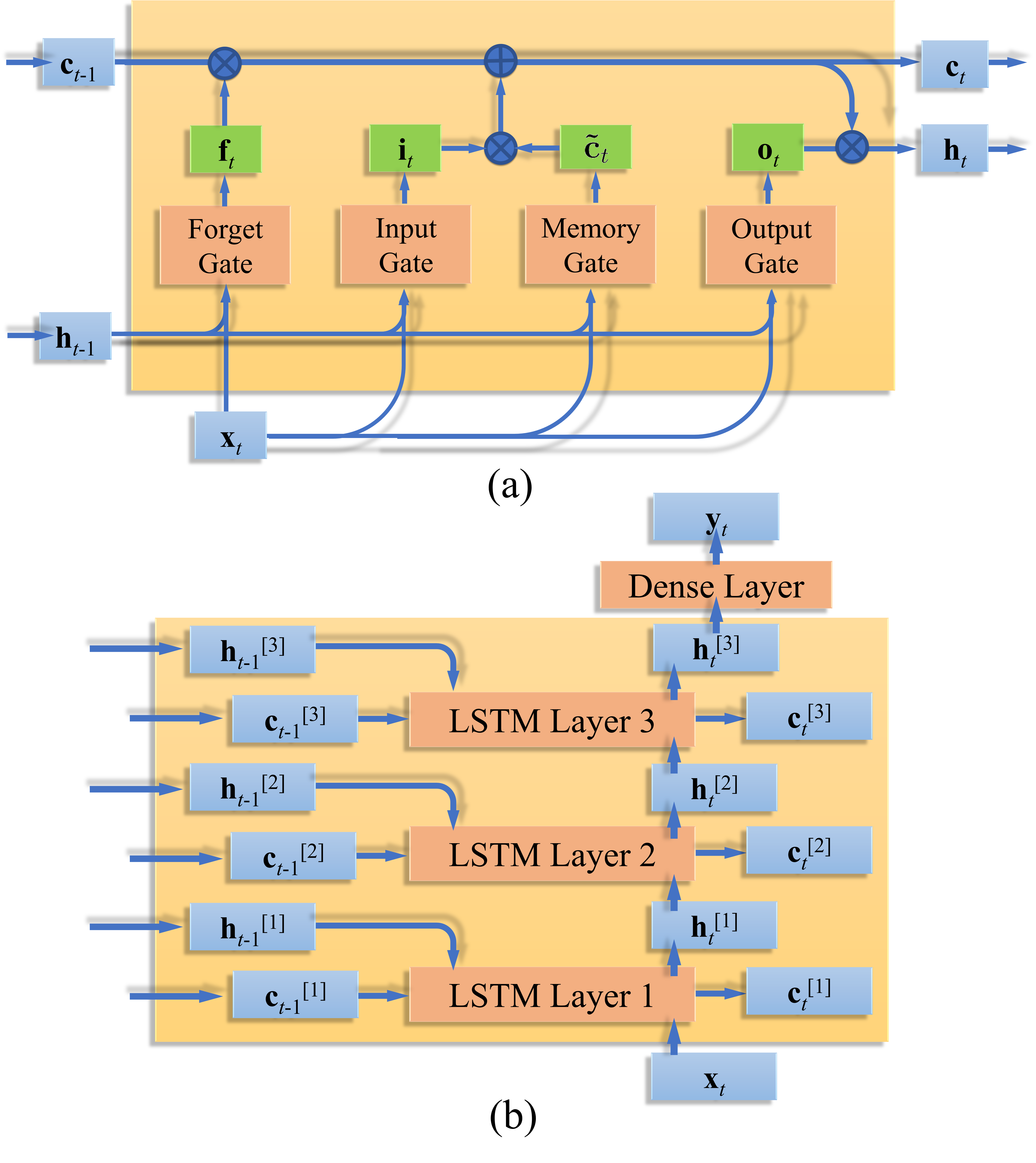}\caption{\label{fig:LSTM} Structures of LSTM unit (a) and multi-layer LSTM
networks (b).}
\end{figure}

In this Letter, we will show that LSTM networks can be incorporated
into FSSH to evolve the electronic density matrix. Let us first give
a brief review about LSTM. Neural network models represent arbitrary
functions with highly interconnected nodes processing through one
or several hidden layers. The input vector $\mathbf{x}$ of the function
is provided to the nodes in input layer, and the output vector $\mathbf{y}$
is obtained from the nodes in output layer. For a conventional $L$-layer
neural network, the connection between two adjacent layers can be
expressed as

\begin{eqnarray}
\mathbf{h^{\mathrm{[\mathit{l}]}}} & = & f\left(\mathbf{W^{\mathrm{[\mathit{l}]}}}\mathbf{h^{\mathrm{[\mathrm{\mathit{l}-1}]}}}+\mathbf{B^{\mathrm{[\mathit{l}]}}}\right)\label{eq:NN}
\end{eqnarray}
where $l=1,2,...,L$. In the above equation, the $l$-th layer receives
$\mathbf{h^{\mathrm{[\mathrm{\mathit{l}-1}]}}}$ from the previous
layer and sends $\mathbf{\mathbf{h^{\mathrm{[\mathit{l}]}}}}$ to
the next layer, $\mathbf{\mathbf{W^{\mathrm{[\mathit{l}]}}}}$ and
$\mathbf{\mathbf{B^{\mathrm{[\mathit{l}]}}}}$ denote weight and bias
parameters for the current layer, respectively, and $f$ is the activation
function such as a sigmoid or hyperbolic tangent function. Note that
$\mathbf{h^{\mathrm{[\mathrm{0}]}}}=\mathbf{x}$ for input layer and
$\mathbf{h^{\mathrm{[\mathrm{\mathit{L}}]}}}=\mathbf{y}$ for output
layer. Recurrent neural network takes this form in general but accepts
a sequence as input variables. A time series $\mathbf{x}_{1},\mathbf{x}_{2},...,\mathbf{x}_{t}$
is a typical input sequence, in which $\mathbf{x}_{t+1}$ at the next
time step can be predicted using neural network as

\begin{eqnarray}
\mathbf{h}_{t} & = & f_{1}\left(\mathbf{W}_{hx}\mathbf{x}_{t}+\mathbf{W}_{hh}\mathbf{h}_{t-1}+\mathbf{B}_{h}\right)\label{eq:RNNh}\\
\mathbf{y}_{t} & = & f_{2}\left(\mathbf{W}_{yh}\mathbf{h}_{t}+\mathbf{B}_{y}\right)\label{eq:RNN}
\end{eqnarray}
where $\mathbf{W}_{hx}$, $\mathbf{W}_{hh}$ and $\mathbf{W}_{yh}$
are weight parameters, $\mathbf{B}_{h}$ and $\mathbf{B}_{y}$ are
bias parameters, $f_{1}$ and $f_{2}$ can be selected as the same
or different activation functions, $\mathbf{h}_{t}$ is the hidden
state vector that implicitly involves historical information of $\mathbf{x}$,
i.e., $\mathbf{x}_{1},\mathbf{x}_{2},...,\mathbf{x}_{t-1}$ through
$\mathbf{h}_{t-1}$ (see Eq \ref{eq:RNNh}), and $\mathbf{x}_{t+1}$
is identical to or obtained based on the output of neural network
($\mathbf{y}_{t}$). The hidden state and all input vectors at previous
time steps are fully-connected in RNN and prone to an exploding or
vanishing gradient. The former can be solved using gradient clipping,
while the latter can be addressed using a gating mechanism such as
long short-term memory networks or gate recurrent unit (GRU). In LSTM,
$\mathbf{h}_{t}$ is employed to represent the short-term state, and
an additional vector $\mathbf{c}_{t}$ is introduced to preserve the
long-term state, which is so-called memory. As shown in Figure \ref{fig:LSTM}(a),
a gated cell is designed to decide what to store by a forget gate
$\mathbf{f}_{t}$ , what to read by an input gate $\mathbf{i}_{t}$,
and what to write by an output gate $\mathbf{o}_{t}$; that is
\begin{eqnarray}
\mathbf{f}_{t} & = & \sigma\left(\mathbf{W}_{fx}\mathbf{x}_{t}+\mathbf{W}_{fh}\mathbf{h}_{t-1}+\mathbf{B}_{f}\right)\label{eq:LSTMf}\\
\mathbf{i}_{t} & = & \sigma\left(\mathbf{W}_{ix}\mathbf{x}_{t}+\mathbf{W}_{ih}\mathbf{h}_{t-1}+\mathbf{B}_{i}\right)\label{eq:LSTMi}\\
\mathbf{o}_{t} & = & \sigma\left(\mathbf{W}_{ox}\mathbf{x}_{t}+\mathbf{W}_{oh}\mathbf{h}_{t-1}+\mathbf{B}_{o}\right)\label{eq:LSTMo}
\end{eqnarray}
where $\sigma$ denotes the sigmoid function. The propagation of LSTM
is given by

\begin{eqnarray}
\mathbf{\tilde{c}}_{t} & = & \tanh\left(\mathbf{W}_{cx}\mathbf{x}_{t}+\mathbf{W}_{ch}\mathbf{h}_{t-1}+\mathbf{B}_{c}\right)\label{eq:LSTMct}\\
\mathbf{c}_{t} & = & \mathbf{f}_{t}\varodot\mathbf{c}_{t-1}+\mathbf{i}_{t}\varodot\mathbf{\mathbf{\tilde{c}}}_{t}\label{eq:LSTMc}\\
\mathbf{h}_{t} & = & \tanh\left(\mathbf{c}_{t}\right)\varodot\mathbf{o}_{t}\label{eq:LSTMh}
\end{eqnarray}
Here $\varodot$ denotes the elementwise product, $\mathbf{y}_{t}$
is obtained from $\mathbf{h}_{t}$ via a dense layer using Eq \ref{eq:RNN}
with a hyperbolic tangent function, and $\mathbf{x}_{t+1}$ is identical
to or obtained based on $\mathbf{y}_{t}$. Eqs \ref{eq:LSTMf}-\ref{eq:LSTMh}
can be briefly expressed as

\begin{eqnarray}
\left\{ \mathbf{h}_{t},\mathbf{c}_{t}\right\}  & = & LSTM\left(\mathbf{x}_{t},\mathbf{h}_{t-1},\mathbf{c}_{t-1};\mathbf{W},\mathbf{B}\right)\label{eq:LSTMch}
\end{eqnarray}
In this work, a multi-layer LSTM is employed and shown in Figure \ref{fig:LSTM}(b);
that is

\begin{eqnarray}
\left\{ \mathbf{h}_{t}^{[l]},\mathbf{c}_{t}^{[l]}\right\}  & = & LSTM\left(\mathbf{h}_{t}^{[l-1]},\mathbf{h}_{t-1}^{[l]},\mathbf{c}_{t-1}^{[l]};\mathbf{W^{\mathrm{[\mathit{l}]}}},\mathbf{\mathbf{B^{\mathrm{[\mathit{l}]}}}}\right)\label{eq:LSTM}
\end{eqnarray}
where $l=1,2,3,$ $\mathbf{h}_{t}^{[0]}=\mathbf{x}_{t}$, and $\mathbf{h}_{t}^{[3]}$
is applied to the dense layer to predict $\mathbf{x}_{t+1}$.

\begin{figure}
\noindent \centering{}\includegraphics[width=0.7\textwidth]{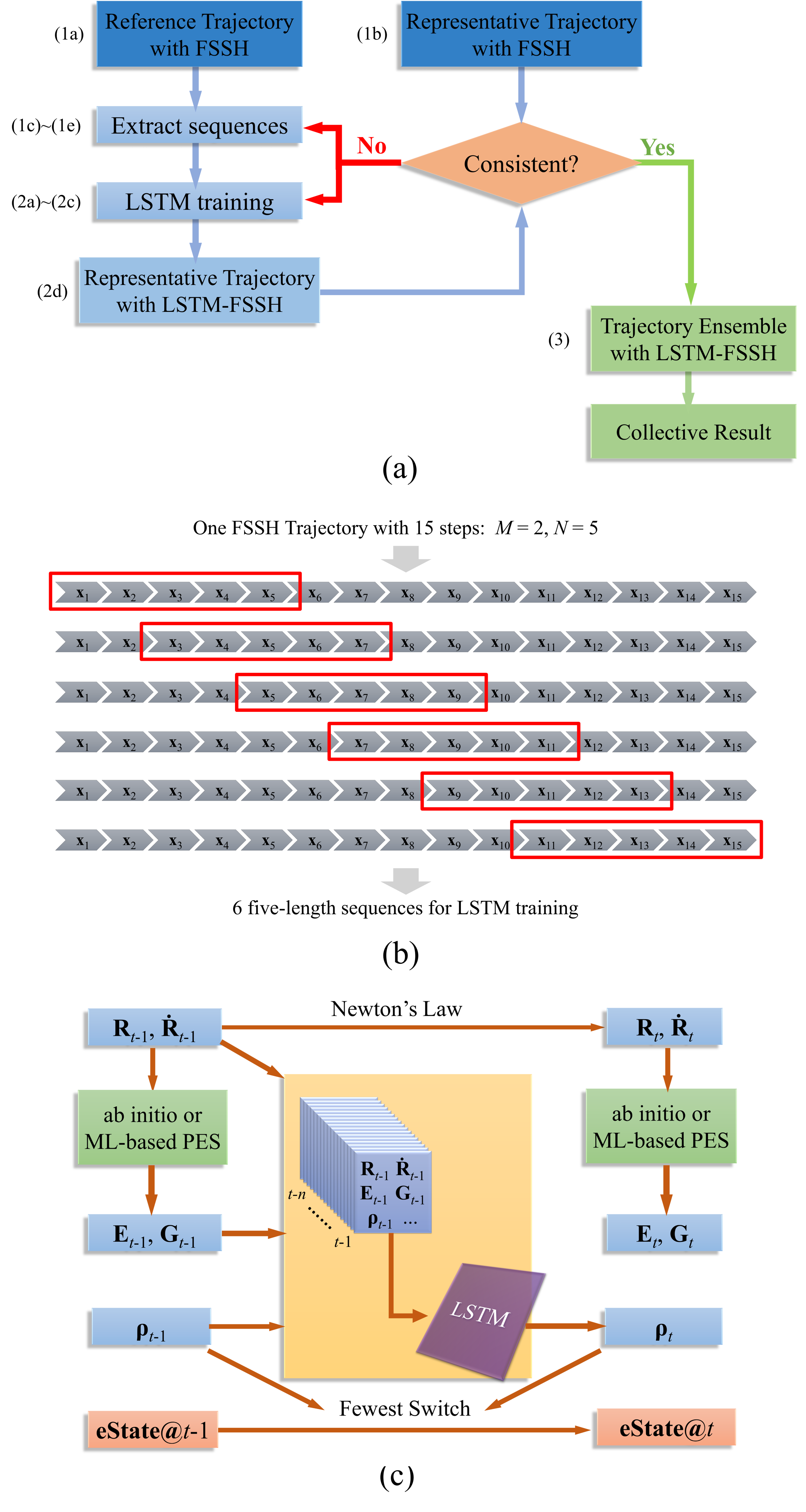}\caption{\label{fig:workflow} Workflow for LSTM-FSSH: whole procedure with
three steps (top), example on generation of sequences from a reference
trajectory in step 1 (middle) and implementation on dynamical simulations
in step 3 (bottom).}
\end{figure}

Here we implemented LSTM networks into the original version of FSSH.
The extension to a variety of modified surface hopping algorithms
is straightforward in principle. In the surface hopping scheme, the
nuclear motion is determined by the gradients on a single adiabatic
PES; that is
\begin{equation}
M_{a}\mathbf{\ddot{R}}_{a}=-\nabla_{\mathbf{R}_{a}}E_{k}\label{eq:EOMn}
\end{equation}
where $\mathbf{R}_{a}$ and $M_{a}$ denote the position and mass
of nucleus $a$, respectively, and $k$ denotes the adiabatic electronic
state at the current time step with the relevant potential energy
$E_{k}$. The time evolution on electronic subsystem can be expressed
as

\begin{equation}
\dot{\rho}_{jk}=-\frac{i}{\hbar}\rho_{jk}\left(E_{j}-E_{k}\right)-\sum_{l}\dot{\mathbf{R}}\cdot\left(\mathbf{d}_{jl}\rho_{lk}-\rho_{jl}\mathbf{d}_{lk}\right)\label{eq:EOMe}
\end{equation}
where $\mathbf{d}_{jk}$ is the nonadiabatic coupling vector between
state $j$ and $k$, and $\mathbf{\rho}$ denotes the electronic density
matrix. According to the assumption of the fewest number of switches,
the probability of nonadiabatic transitions from the current state
$k$ to state $j$ in a time interval $\Delta t$ is

\begin{equation}
P_{k\rightarrow j}\left(t,t+\Delta t\right)=\max\left(0,\frac{\int_{t}^{t+\Delta t}2\mathrm{Re}\left(\dot{\mathbf{R}}\cdot\mathbf{d}_{kj}\rho_{jk}\right)d\tau}{\rho_{kk}\left(t\right)}\right)\label{eq:Prob}
\end{equation}
For a system only involving two electronic states, the probability
of switching from the current state $k$ to state $j$ is identical
to that of jumping out from state $k$, which can be expressed as\cite{Tully_1990}

\begin{equation}
P_{k\rightarrow j}\left(t,t+\Delta t\right)=\max\left(0,\frac{\rho_{kk}\left(t\right)-\rho_{kk}\left(t+\Delta t\right)}{\rho_{kk}\left(t\right)}\right)\label{eq:Probml}
\end{equation}
In presence of multiple electronic states, Eq \ref{eq:Probml} is
the summation over hopping probabilities from the current state to
all other states and cannot be used directly. During FSSH simulations
on molecular systems, however, the hopping is only allowed when the
energy gap is lower than a certain threshold (e.g., 10 kcal/mol) in
practice. It means that in most regions of PES, the quantum subsystem
can be reduced to the simple two-state case. It is no longer reliable
in the vicinity of multistate intersections, in which some advanced
nonadiabatic dynamical simulations beyond FSSH are usually required.

During conventional FSSH simulations, Eqs \ref{eq:EOMn} and \ref{eq:EOMe}
are used to govern nuclear and electronic motion, respectively, and
the hopping probability is calculated with Eq \ref{eq:Prob}. In LSTM-FSSH,
a LSTM network is built based on a small number of existing trajectories
and works as a propagator of electronic degrees of freedom, which
replaces Eq \ref{eq:EOMe}. In other words, $\mathbf{\mathbf{\rho}}\left(\mathit{t}\mathrm{+\Delta}\mathit{t}\right)$
is predicted by LSTM based on a time series in the preceding trajectory
without solving Eq \ref{eq:EOMe}. The hopping probability is obtained
directly using Eq \ref{eq:Probml} instead of Eq \ref{eq:Prob}. On
one hand, the time evolution on classical subsystem can be also learned
by LSTM or RNN as reported in recent works.\cite{Wang_2020,Vlachas_2021}
On the other hand, several ML-based protocols have been performed
excellently to predict accurate potential energies and gradients on
the right-hand side of Eq \ref{eq:EOMn}, which may be a better choice
for realistic molecular systems.\cite{Behler_2017,Pinheiro_2021}
Here the classical subsystem still evolves according to Eq \ref{eq:EOMn}
with analytical gradients of our test model systems. More benchmarks
and further methodology developments on this issue are left for future
work.

The workflow of LSTM-FSSH is summarized in Figure \ref{fig:workflow}.
We take a one-dimensional system with two electronic states as an
instance and outline the procedure as follows:

Step 1. Database construction.

(1a) Conventional FSSH simulations with a time step $\mathrm{\Delta}\mathit{t}$
are performed to generate a small number of reference trajectories.
The nuclear position $\mathbf{R}$ and velocity $\mathbf{\dot{R}}$,
potential energies and gradients in two states as $E_{0}$, $E_{1}$,
$\mathbf{G}_{0}$, $\mathbf{G}_{1}$, as well as two independent elements
of electronic density matrix as $\rho_{00}$ and $\rho_{01}$, are
recorded at each step.

(1b) Select a few representative trajectories as ``external reference
trajectories'' in preparation for (2d). For example, three trajectories
at a low, medium and high initial momentum for each model are selected
in this work. These trajectories are excluded from reference trajectories
in (1c)-(2c).

(1c) A time series $\mathbf{x}_{1},\mathbf{x}_{2},...,\mathbf{x}_{N}$
with $N$ frames is extracted with a time step $\mathrm{\Delta}\mathit{t}_{s}$
from one reference trajectory. Here $\mathbf{x}$ consists of $\mathbf{R}$,
$\mathbf{\dot{R}}$, $E_{0}$, $E_{1}$, $\mathbf{G}_{0}$, $\mathbf{G}_{1}$,
$\rho_{00}$, $\rho_{01}$ and/or arbitrary functions of the above
nuclear degrees of freedom, and $\mathrm{\Delta}t_{s}$ is equal to
$\mathrm{\Delta}\mathit{t}$ for simplicity.

(1d) The sequence obtained in (1c) is shifted forward by $M$ frames
with the same time step to extract another time series until the end
of this reference trajectory.

(1e) Repeat (1c) and (1d) on other reference trajectories to extract
a large number of $N$-length sequences as the database for LSTM training.

Step 2. LSTM training.

(2a) Select 80\% of reference sequences randomly from the database
to build the training set. The remaining sequences belong to the testing
set.

(2b) Set initial values of hyperparameters of LSTM, such as layer
sizes of network and batch sizes for minimization.

(2c) Perform the training of LSTM based on the training set. The mean
squared error (MSE) of $\rho_{00}$ and $\rho_{01}$ is minimized
using Adam optimizer\cite{Kingma_2014}, and the MSE for the testing
set is monitored to avoid overfitting.

(2d) Perform LSTM-FSSH with the same initial condition and random
numbers for hopping of the external reference trajectories selected
in (1b). If the generated trajectory visually agrees with the corresponding
FSSH trajectory, the present LSTM model is acceptable. Otherwise,
return to (2c) to rebuild LSTM using different hyperparameters and
random numbers for training.

Step 3. FSSH simulation.

(3a) Initialize the nuclear position, velocity and electronic density
matrix for one trajectory as the same as conventional FSSH simulation
without machine learning.

(3b) Perform conventional FSSH simulation using Eqs \ref{eq:EOMn},
\ref{eq:EOMe} and \ref{eq:Prob} for $N$ steps to produce an $N$-length
sequence, that is, a short-time dynamics in absence of LSTM in the
beginning.

(3c) The generated sequence is provided into LSTM to predict electronic
density matrix in the next step. Note that the dimensionalities of
input and output features of LSTM can be different. Take our test
case as an instance. The output layer usually has three nodes as $\rho_{00}$,
$\mathrm{Re\left(\rho_{01}\right)}$ and $\mathrm{Im\left(\rho_{01}\right)}$,
while the input layer includes all nuclear and electronic features.

(3d) The nuclear position and velocity are propagated to the next
step using Eq \ref{eq:EOMn}, followed by the calculation on potential
energies and gradients in all electronic states. The hopping probability
between electronic states is obtained using Eq \ref{eq:Probml}, and
the nuclear velocity is adjusted if a switch occurs. This step is
the same as conventional FSSH simulation without machine learning,
except for the calculation on hopping probability.

(3e) Shift the $N$-length sequence forward by one step. Now the electronic
and nuclear features updated in (3c) and (3d) is added into the sequence.

(3f) Return to (3c) to propagate the trajectory until the stopping
criterion such as the maximum MD step is achieved.

(3g) Repeat (3a)-(3f) to produce a large number of trajectories. Study
collective photodynamic behavior of the system based on trajectory
ensemble.

It is worthy noting that step (2d) is the most essential during the
whole procedure. Unlike the common case of ML training, step (2c)
converges very quickly and seems to be insensitive to hyperparameters
in our test case. It indicates that the MSE for the testing set may
be insufficient to reflect actual performance of LSTM on dynamical
simulations. Actually, it is observed that a set of LSTM networks,
all of which perform excellently in (2c), usually produce diverse
FSSH trajectories with the same initial condition and random numbers.
It may be due to the cumulative error of ML propagator after a long-time
ML-driven dynamical simulation. The implementation in step (2d) is
thus required. In this work, we compare the time evolution of electronic
density matrix visually as a criterion of consistency between external
reference and LSTM-FSSH trajectories. The hyperparameters $M$ and
$N$ in step 1 are also tuned, leading to a reconstructed database.
In practice, the input features as well as the arbitrary functions
mentioned in step (1c) may be even necessary to select again before
returning to (2c). However, the additional computational cost is small
because the reference FSSH trajectories have been produced in step
(1a) and keep the same regardless of database reconstruction or input
feature reselection.

\begin{figure}
\noindent \centering{}\includegraphics[width=1.05\textwidth]{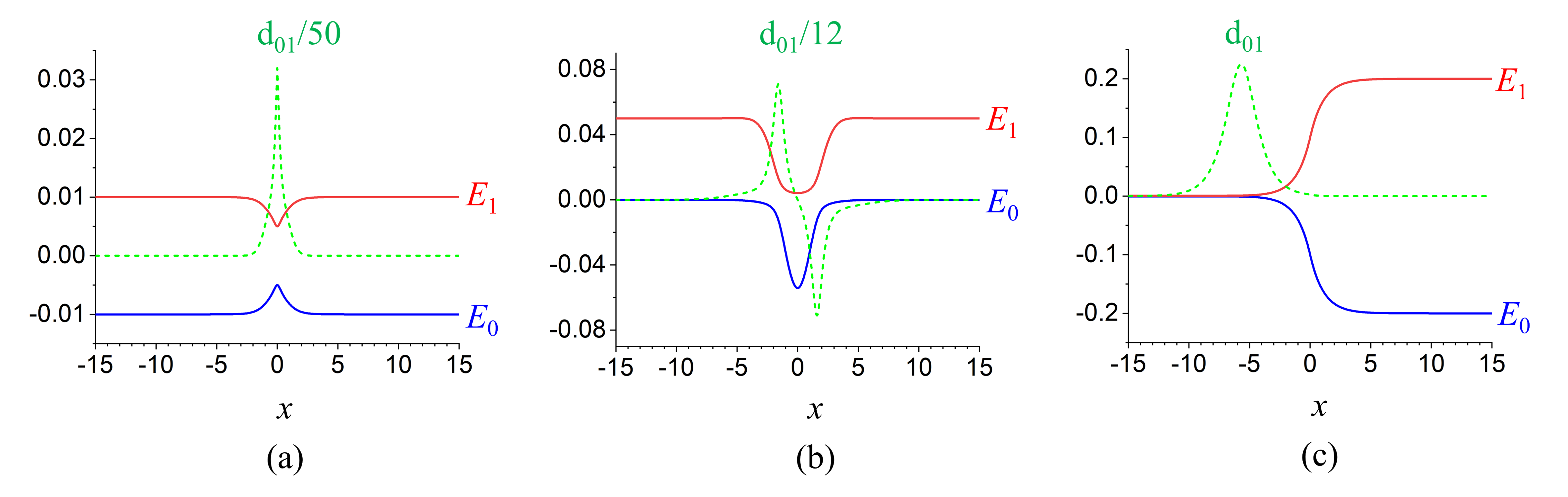}\caption{\label{fig:model} Adiabatic PESs and nonadiabatic coupling vectors
of the single-avoided crossing model (a), the dual-avoided crossing
model (b), and the extended coupling Hamiltonian model (c). All quantities
are in atomic units.}
\end{figure}

\begin{figure}
\noindent \centering{}\includegraphics[width=1\textwidth]{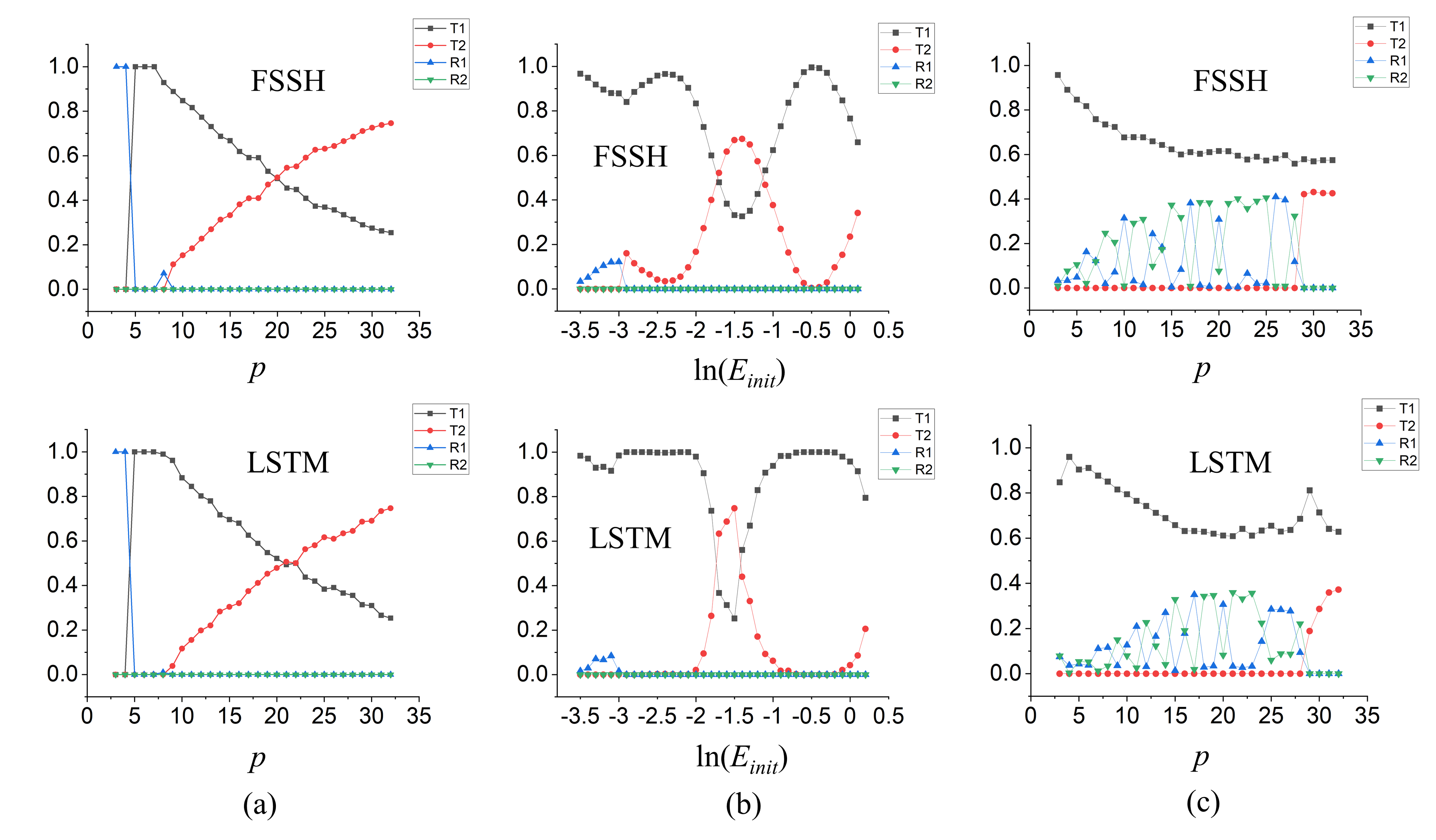}\caption{\label{fig:collect} Collective results of single-avoided crossing
(a), dual-avoided crossing (b), and extended coupling models (c) obtained
using original FSSH as reference (top) or LSTM-FSSH (bottom).}
\end{figure}

We employ Tully’s three models reported in the original FSSH paper\cite{Tully_1990}:
the single-avoided crossing model, the dual-avoided crossing model,
and the extended coupling model as our test systems (Fig. \ref{fig:model}),
which are supposed to cover most of the typical features in realistic
molecular systems. The reference trajectories as well as the final
collective results for comparison are obtained using the original
FSSH method despite its well-known limitations such as the overcoherence
problem.\cite{Nelson_2013,Subotnik_2016,Plasser_2019,Tang_2021} All
parameters of Tully's models are in atomic units. In order to keep
consistent with previous works, the classical position $\mathbf{R}$
and velocity $\mathbf{\dot{R}}$ in Fig. \ref{fig:workflow}(c) are
represented by $x$ and $v$, respectively, for Tully's three models.
All trajectories start at $x=-20.0$ on the lower surface with an
initial momentum to the right and stop at $x=\pm25.0$. The elements
of electronic density matrix are initialized as zero except $\rho_{00}=1.0$.
The time step $\mathrm{\Delta}\mathit{t}=1.0$ for all simulations,
and $\mathrm{\Delta}t_{s}$ for LSTM is equal to $\mathrm{\Delta}\mathit{t}$.
The mass in Eq \ref{eq:EOMn} is set as 2000 to mimic atomic nuclei.
For each model and each method, 2000 trajectories are simulated with
each initial momentum for collection. The collective results consist
of four channels: transmission to $x>25.0$ in the lower state (T1),
transmission to $x>25.0$ in the upper state (T2), reflection to $x<-25.0$
in the lower state (R1) and reflection to $x<-25.0$ in the upper
state (R2). The proportions that the trajectory stops in each channel
are calculated and shown in Fig. \ref{fig:collect}. The training
of all LSTM networks are implemented using Keras\cite{Keras} combined
with Tensorflow\cite{Tensorflow}. The interface between FSSH propagator
and LSTM networks are built using Keras2C\cite{Keras2c}.

The diabatic Hamiltonian of the single-avoided crossing model is given
as

\begin{eqnarray}
V_{11}\left(x\right) & = & \begin{cases}
A\left[1-e^{-Bx}\right] & x>0\\
-A\left[1-e^{Bx}\right] & x<0
\end{cases}\nonumber \\
V_{22}\left(x\right) & = & -V_{11}\left(x\right)\label{eq:Tully1}\\
V_{12}\left(x\right) & = & V_{21}\left(x\right)=Ce^{-Dx^{2}}\nonumber 
\end{eqnarray}
where $A=0.01$, $B=1.6$, $C=0.005$, and $D=1.0$. The corresponding
adiabatic PESs and NACVs are shown in Figure \ref{fig:model}(a).
We generate 20 reference trajectories without machine learning for
each initial momentum. The sequences in the training and testing sets
are extracted from all reference trajectories with all initial conditions.
The classical position $x$, velocity $v$, potential energies $E_{0}$
and $E_{1}$, as well as elements of electronic density matrix are
employed as input features. More details can be seen in Tables S1,
S2 and S3.

\begin{figure}
\noindent \centering{}\includegraphics[width=1\textwidth]{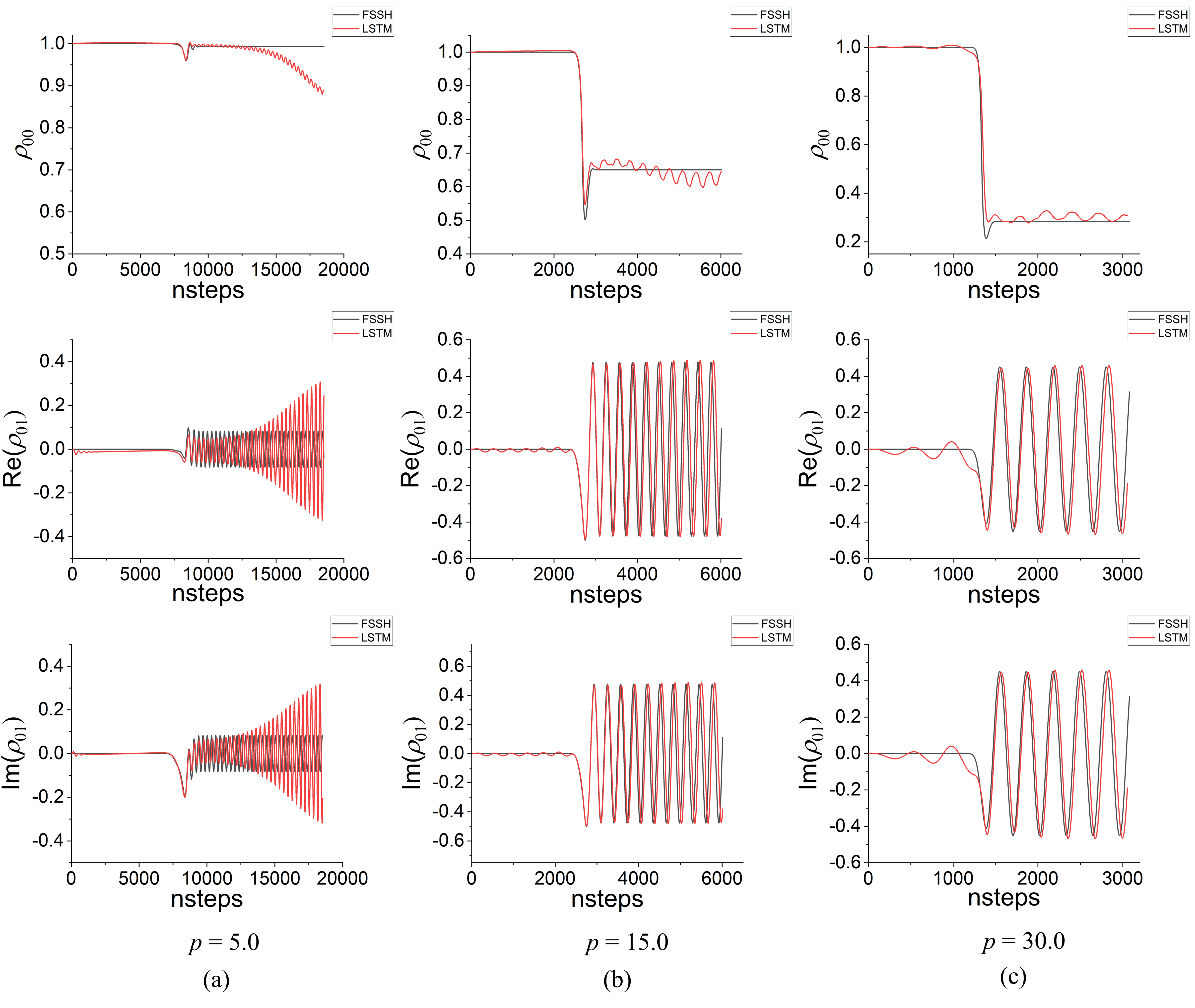}\caption{\label{fig:Tully1} Electronic density matrix plotted as function
of time for representative trajectories of single-avoided crossing
model at a low (a), medium (b) and high (c) initial momentum. Different
colors represent different simulation methods (black: original FSSH
as reference; red: LSTM-FSSH).}
\end{figure}

The excellent performance for the testing set in step (2c) is shown
in Figure S1. As mentioned in step (2d), LSTM-FSSH simulations are
implemented to produce three representative trajectories with different
initial momenta. The time evolution of $\rho_{00}$, $\mathrm{Re\left(\rho_{01}\right)}$
and $\mathrm{Im\left(\rho_{01}\right)}$ is compared to the corresponding
reference trajectories with the same initial condition and random
numbers for hopping (see Figure \ref{fig:Tully1}). At a low momentum
of 5.0, LSTM-FSSH keeps consistent with the reference for 12500 steps.
The deviation takes place after the system departs from the coupling
region, which will not affect collective results. At a medium momentum
of 15.0 and a large momentum of 30.0, LSTM-FSSH agrees well with the
reference except for a slightly fluctuation of $\rho_{00}$ after
a sudden decay. The hopping event is observed only at a large momentum.
The system switches to the upper surface at step 1882 and 1339 with
and without LSTM, respectively. Then we run 2000 trajectories with
each initial momentum (60000 in total) using LSTM-FSSH. As shown in
Figure \ref{fig:collect}, the collective result is almost the same
as the original FSSH simulations. The resonance phenomenon at $7.0<p<9.0$
populated on R1 can be partially captured by machine learning with
a smaller probability.

The diabatic Hamiltonian of the dual-avoided crossing model is
\begin{eqnarray}
V_{11}\left(x\right) & = & 0\nonumber \\
V_{22}\left(x\right) & = & -Ae^{-Bx^{2}}+E\label{eq:Tully2}\\
V_{12}\left(x\right) & = & V_{21}\left(x\right)=Ce^{-Dx^{2}}\nonumber 
\end{eqnarray}
where $A=0.10$, $B=0.28$, $C=0.015$, $D=0.06$, and $E=0.05$.
The adiabatic PESs and NACVs are shown in Figure \ref{fig:model}(b),
where $E_{init}$ denotes the initial kinetic energy, and the initial
momentum $p=\sqrt{2mE_{init}}$. For each initial momentum, there
are 50 reference trajectories generated without machine learning.
Then the reference sequences are extracted to build the database for
LSTM training, covering all initial momenta. The input features are
selected as the same as the first model. More details can be seen
in Tables S1, S2 and S4.

\begin{figure}
\noindent \centering{}\includegraphics[width=1\textwidth]{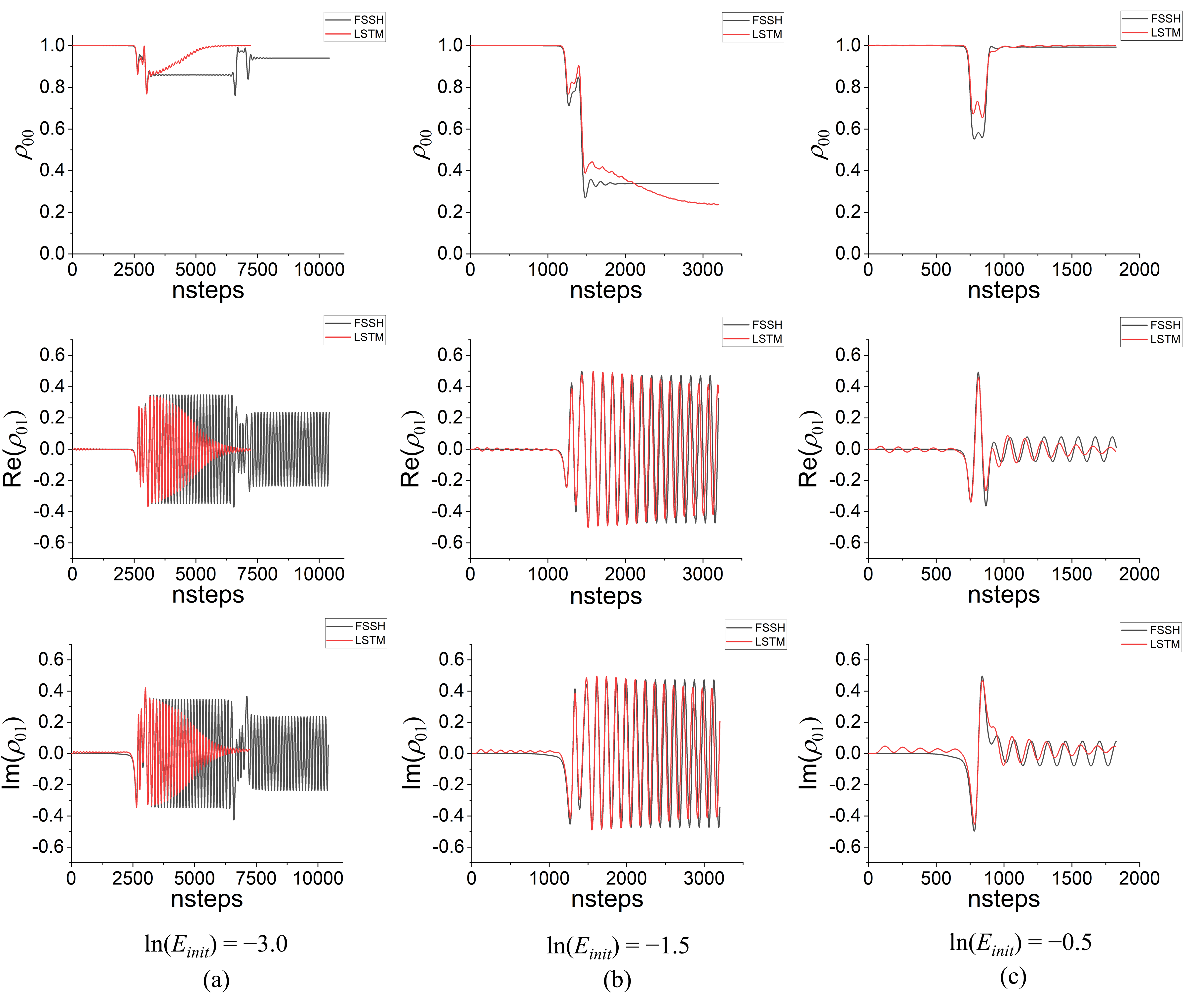}\caption{\label{fig:Tully2} Electronic density matrix plotted as function
of time for representative trajectories of dual-avoided crossing model
at a low (a), medium (b) and high (c) initial momentum. Different
colors represent different simulation methods (black: original FSSH
as reference; red: LSTM-FSSH).}
\end{figure}

First, it can be seen from Figure S1 that the performance for the
testing set is remarkable. Second, three representative trajectories
using LSTM-FSSH simulations are compared. As shown in Figure \ref{fig:Tully2},
the higher the initial momentum, the better the performance of machine
learning. At a low momentum of $\ln(E_{init})=-3.0$, LSTM-FSSH deviates
away from the reference after step 3000. Both trajectories switch
to the upper surface at step 2949, and return to the lower surface
at step 4257 and 7156 using LSTM and FSSH, respectively. At a medium
momentum of $\ln(E_{init})=-1.5$, two trajectories share the same
hopping event at step 1424. At a high momentum of $\ln(E_{init})=-0.5$,
the transition between two surfaces is never observed in the representative
trajectories using LSTM, while the original FSSH simulation predicts
an excited-state population lasting for a short time around step 800.
Finally, 2000 LSTM-FSSH trajectories with each initial momentum (74000
in total) are implemented. The collective result shown in Figure \ref{fig:collect}
is consistent with the performance on representative trajectories.
Our method gives a qualitatively correct result except an error within
$-3.0<\ln(E_{init})<-2.0$ where the increase of T2 population is
missing. It is probably due to the relatively small number of T2 trajectories
in our training set, but manually picking trajectories with a specified
population is always unrealistic for polyatomic molecules.

The simulation results using another LSTM model with different hyperparameters
are reported in Figs. S1, S17 and S18. No problem can be observed
from the MSE of the testing set in step (2c). However, the FSSH simulations
with and without machine learning in step (2d) represent opposite
tendencies of the time evolution of $\rho_{00}$. At a low momentum,
$\rho_{00}$ keeps around 1.0 in the reference but decreases to 0.5
using LSTM after a long-time dynamics. At a medium momentum, $\rho_{00}$
converges to 0.3 in the reference but returns to 0.8 using LSTM. It
indicates large cumulative errors of this inappropriate LSTM model,
leading to an incorrect collective result.

The diabatic Hamiltonian of the extended coupling model is
\begin{eqnarray}
V_{11}\left(x\right) & = & A\nonumber \\
V_{22}\left(x\right) & = & -A\label{eq:Tully3}\\
V_{12}\left(x\right) & = & V_{21}\left(x\right)=\begin{cases}
B\left[2-e^{-Cx}\right] & x>0\\
Be^{Cx} & x<0
\end{cases}\nonumber 
\end{eqnarray}
where $A=0.0006$, $B=0.1$, and $C=0.9$. The adiabatic PESs and
NACVs are shown in Figure \ref{fig:model}(c). Note again that the
original FSSH simulation results are employed as the reference of
machine learning, and the decoherence correction, which is necessary
for this model to reproduce its full-quantum dynamics behavior, is
out of our scope in this work. We first implement FSSH simulations
without machine learning to produce 50 reference trajectories at each
initial momentum. However, all attempts on LSTM training using the
above strategy are unsuccessful in step (2d). As shown in Fig. S19,
the decrease of $\rho_{00}$ appears too early using LSTM; LSTM also
fails to reproduce the strong oscillation of $\rho_{01}$, which is
relevant to the coherence in the coupling region. Unsurprisingly,
the collective result deviates away from the reference (see Fig. S20).

We refine the LSTM network in two ways. First, since the coupling
between quantum and classical subsystems is the key factor of MQC-MD,
elaborate selection on input features related to nuclear degrees of
freedom is expected to improve the performance of LSTM on the prediction
of electronic motion. Functions of nuclear degrees of freedom: $\frac{G_{0}\left|v\right|}{\left(E_{1}-E_{0}\right)^{2}}$
and $\frac{G_{1}\left|v\right|}{\left(E_{1}-E_{0}\right)^{2}}$, are
introduced as two additional input features. Three factors including
the degeneracy of PESs, the branching of PESs and the correlation
between nuclear velocity and transition probability in the intersection
region are considered. It would be beneficial to add a small positive
bias $\varepsilon$ to denominator when the potential energy difference
is very close to zero, but we have never encountered such a problem
on numerical instability in the present work. Second, multiple connected
LSTM networks are constructed. As shown in Fig. S11(a-d), the first
LSTM network is used at the beginning of trajectory ($x=-20.0$).
When the system first reaches $x=-2.0$, the second LSTM is employed
until it first reaches $x=1.0$ or returns to $-2.0$. The third LSTM
is applied sequentially until the system first reaches $x=4.0$ or
returns to $-7.0$. The last LSTM is responsible when $x\geq4.0$
or $\leq-7.0$. Two infrequent cases were also displayed in Fig. S11(e-f),
in which the velocity reverses when $x\leq-2.0$ or $\geq4.0$ at
a very low or high momentum, respectively. In the former case, the
first LSTM would switch to the third one immediately; in the latter
case, the last LSTM would change to the third one when the system
returns to $x=4.0$. The third network would be kept until $x\leq-7.0$
in both cases, followed by the last LSTM. More details of LSTM used
for the extended coupling model can be seen in Tables S1, S2 and S5.

\begin{figure}
\noindent \centering{}\includegraphics[width=1\textwidth]{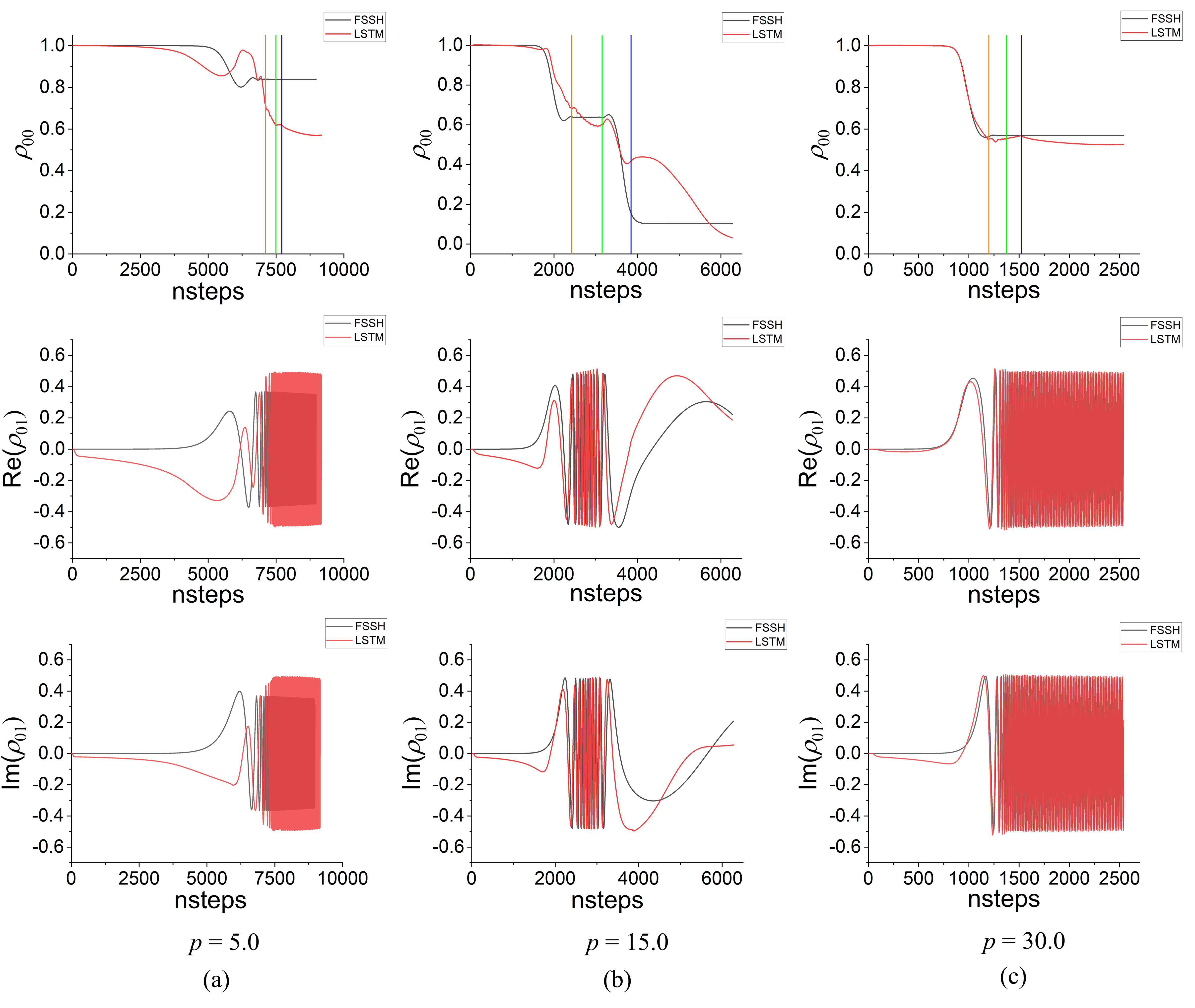}\caption{\label{fig:Tully3} Electronic density matrix plotted as function
of time for representative trajectories of extended coupling model
at a low (a), medium (b) and high (c) initial momentum. Different
colors represent different simulation methods (black: original FSSH
as reference; red: LSTM-FSSH). Switches between two connected LSTM
networks are represented by vertical lines (orange: $1\mathrm{\rightarrow}2$;
green: $2\mathrm{\rightarrow}3$; blue: $3\mathrm{\rightarrow}4$).}
\end{figure}

The comparison between three representative LSTM-FSSH trajectories
and their corresponding references is shown in Figure \ref{fig:Tully3}.
The multiple networks perform well. The time evolution of $\rho_{00}$,
$\mathrm{Re}(\rho_{01})$ and $\mathrm{Im}(\rho_{01})$ agrees with
the reference except for a problem at a low momentum of 5.0, where
$\rho_{00}$ descends too early and tends to oscillate using LSTM.
Hopping events are observed only at a medium momentum of 15.0 with
both methods, taking place at the same step as 1927. The collective
result shown in Figure \ref{fig:collect} is obtained from 2000 LSTM-FSSH
trajectories with each initial momentum (60000 in total). LSTM-FSSH
is qualitatively identical to the original FSSH simulations without
machine learning. Further analysis on the distribution of hopping
events is shown in Figures S9 and S10, disclosing a few unphysical
predictions of LSTM after long-time dynamics. This can be attributed
to not only the cumulative errors of ML propagator but also the tendency
of LSTM-FSSH to learn the average features of trajectory ensemble,
in which trajectories starting from the same initial condition should
give different time series due to random hopping events. Despite these
intrinsic limitations of our method, their influence on simulation
results is possible to evaluate or control at least roughly (see Section
S2.3, SI). In addition, the MSEs for the testing sets with all LSTM
models are indistinguishable in Figure S1 despite different collective
results, which highlights the importance of step (2d) on LSTM training.
Using a single LSTM with additional input features only partially
improves the performance of collective results (see Section S3.2,
SI).

Although the improvement of multiple LSTM networks with additional
input features is remarkable, we still suggest simple settings of
LSTM networks as the first choice on a new system. First, nuclear
positions and velocity, potential energies in two electronic states,
and elements of electronic density matrix are recommended as initial
input features. More complex descriptor such as $\frac{G_{0}\left|v\right|}{\left(E_{1}-E_{0}\right)^{2}}$
and $\frac{G_{1}\left|v\right|}{\left(E_{1}-E_{0}\right)^{2}}$, which
also performs well for other test cases in this work (see Fig. S16),
can be introduced when LSTM network with different hyperparameters
cannot work in step (2d). Second, a close examination on representative
trajectories indicates that the discrepancy between energy degeneration
and strong coupling regions may lead to the difficulties on the extended
coupling model. According to our experience, however, it is an unusual
case for realistic photochemical reactions, which can be identified
by detecting the coupling region based on frequent hopping events
in reference trajectories. Multiple networks provide an option if
all attempts using a single LSTM fail. Different networks can be applied
to the detected strong coupling region, the intersection region without
strong coupling and other regions with a large energy gap between
two states.

The application of this method on realistic molecular systems is attractive,
but some nontrivial issues should be addressed in prior. First, although
NACV is no longer necessary for two-state systems, the excited-state
calculation on potential energies and gradients is still a bottleneck
for medium-sized or larger molecules. Fortunately, ML-based force
fields have been successfully applied to dynamical simulations on
multiple PESs of several organic molecules in photochemistry.\cite{Hu_2018,Chen_2018,Menger_2020,Li_2021,Lin_2021jcp}
The combination of ML-based force fields and LSTM-FSSH is straightforward,
in which the term on the right-hand side of Eq \ref{eq:EOMn} would
be obtained from ML prediction instead of electronic structure calculation.
Second, the input feature selection is more complicated for high-dimensional
classical subsystem. Our preliminary results show that a large number
of input features related to nuclear motion hamper the ability of
LSTM to propagate electronic subsystem. The encoder network that generates
a low-order latent space may be a promising choice to extract more
concentrated input features.\cite{Vlachas_2021} The use of encoder
in LSTM-FSSH would be simpler than latent-space classical dynamics
because the time evolution on nuclei is still implemented in the full-dimensional
space. Third, a general quantitative criterion in step (2d) is absent.
Here we compare the representative LSTM-FSSH trajectories with the
reference visually, which would be system-specific and laborious for
realistic molecules. ML-based analysis of MQC-MD dynamics such as
isometric feature mapping and multidimensional scaling provides a
possible solution to this trouble.\cite{Li_2017} Fourth, the time-series
LSTM predictor always suffers from cumulative errors, while non-time-series
ML approaches, which have been implemented for surface hopping dynamics
simulations on realistic molecular systems, are free of this problem.
More comparisons between LSTM-FSSH and the non-time-series ML-based
excited-state dynamics methods such as SchNarc \cite{Westermayr_2020}
should be performed carefully. Finally, how to extend this method
to study the dynamics around multistate intersections is unsolved
in this work. We suggest making a pause for ML-driven simulation and
switching to a more accurate MQC-MD approach\cite{Crespo_Otero_2018,Curchod_2018,Shen_2019}
when the system evolves in multistate intersection regions.

In summary, we have demonstrated how to construct long short-term
memory networks as a propagator of electronic subsystem in fewest-switches
surface hopping simulations. The single-avoided crossing, dual-avoided
crossing and extended coupling models are employed as our test systems.
Starting with a small number of full-length trajectories (20-50 for
each initial momentum and each model) from the original FSSH simulations,
LSTM networks can be built to reproduce the time evolution of electronic
density matrix during a long-time FSSH dynamical simulation in presence
of only 50-100 steps given as the initial input sequence. Then we
simulate 2000 trajectories for each initial momentum and each model
using LSTM-FSSH. The collective results is qualitatively consistent
with the original FSSH as reference, even though the extended coupling
model is much more difficult than other test cases and requires multiple
connected LSTM networks. All results show that a lower initial momentum
results in a longer-time trajectory with larger cumulative errors,
which is an intrinsic drawback of LSTM propagators. Besides avoiding
expensive computation on nonadiabatic coupling vectors at least for
two-state systems, the combination of LSTM and ML-based force field
has a great potential to further accelerate surface hopping simulations
on realistic molecular systems. We believe that LSTM is a powerful
tool on mixed quantum-classical molecular dynamics to study photophysics
and photochemistry more effectively.

\subsection*{Corresponding Author}

{*}Email: lshen@bnu.edu.cn

\subsection*{Supplementary Information}

Electronic supplementary information (ESI) including input features
and hyperparameters of LSTMs, additional results of LSTM-FSSH, more
information on LSTM networks for extended coupling model, and other
LSTM networks with unsatisfactory performance is available. The source
code is available on https://github.com/TDD365/LSTM-FSSH, under the
GNU General Public License v3.0.

\subsection*{Acknowledgments}

We are grateful for the financial support from the NSFC for L. S.
(Nos. 21903005 and 22193041) and W. F. (No. 21688102) and the Beijing
Normal University Startup for L. S..

\bibliography{manu}


\section*{Supplementary material}

# Supplementary Information for: Fewest-Switches Surface Hopping with Long Short-Term Memory Networks

Diandong Tang,[†] Luyang Jia,[†] Lin Shen,[*,†,‡] and Wei-Hai Fang[†,‡]

[†]Key Laboratory of Theoretical and Computational Photochemistry of Ministry of Education, College of Chemistry, Beijing Normal University, Beijing 100875, China

[‡]Yantai-Jingshi Institute of Material Genome Engineering, Yantai 265505 Shandong, China

E-mail: lshen@bnu.edu.cn

# Table of Contents

# Section S1. Input features and hyperparameters of LSTMs

Table S1. Input features of LSTMs for Tully's three models

| System  |   |   |       |       |             |                |                |                            |                            |
|---------|---|---|-------|-------|-------------|----------------|----------------|----------------------------|----------------------------|
|         |   |   |       |       | Input Feature |              |                |                            |                            |
| Tully 1 | $x$ | $v$ | $E_0$ | $E_1$ | $\rho_{00}$ | $Re(\rho_{01})$ | $Im(\rho_{01})$ |                            |                            |
| Tully 2 | $x$ | $v$ | $E_0$ | $E_1$ | $\rho_{00}$ | $Re(\rho_{01})$ | $Im(\rho_{01})$ |                            |                            |
| Tully 3 | $x$ | $v$ | $E_0$ | $E_1$ | $\rho_{00}$ | $Re(\rho_{01})$ | $Im(\rho_{01})$ | $\frac{G_0|v|}{(E_1-E_0)^2}$ | $\frac{G_1|v|}{(E_1-E_0)^2}$ |

Tully 1: Single-avoided crossing model.
Tully 2: Dual-avoided crossing model.
Tully 3: Extended coupling model.
$x$: classical position.
$v$: classical velocity.
$E_0$, $G_0$: potential energy and gradient in lower surface.
$E_1$, $G_1$: potential energy and gradient in upper surface.
$\rho_{00}$, $\rho_{01}$: elements of electronic density matrix.


Table S2. Hyperparameters of LSTMs for Tully's three models

|                  | Tully 1 | Tully 2 | Tully 3   |           |           |           |
|------------------|---------|---------|-----------|-----------|-----------|-----------|
|                  |         |         | Network 1 | Network 2 | Network 3 | Network 4 |
| $N$_input [a]    | 7       | 7       | 9         | 9         | 9         | 9         |
| Size [b]         | 32      | 48      | 72        | 72        | 72        | 72        |
| $N$ [c]          | 100     | 50      | 50        | 50        | 50        | 50        |
| $M$ [d]          | 50      | 20      | 25        | 25        | 3         | 5         |
| $N$_train [e]    | 96235   | 355776  | 191903    | 35655     | 175480    | 471559    |
| $N$_batch [f]    | 64      | 32      | 32        | 16        | 32        | 64        |
| $\alpha$ [g]     | 0.001   | 0.001   | 0.001     | 0.001     | 0.001     | 0.001     |

[a] Number of input features.
[b] Number of nodes for each LSTM layer and dense layer.
[c] Length of input sequence; see explanation for step (1c) in main text.
[d] Interval between extracted sequences; see explanation for step (1d) in main text.
[e] Number of sequences in training set.
[f] Batch size for minimization.
[g] Learning rate.

Table S3. Scaled input features of LSTM for single-avoided crossing model.

| Definition | Scaled |
|---|---|
| $x$ | $0.05x$ |
| $v$ | $(2000v-17)/30$ |
| $E_0$ | $50E_0$ |
| $E_1$ | $50E_1$ |
| $\rho_{00}$ | $\rho_{00}-0.5$ |
| $\text{Re}(\rho_{01})$ | $\text{Re}(\rho_{01})$ |
| $\text{Im}(\rho_{01})$ | $\text{Im}(\rho_{01})$ |

Table S4. Scaled input features of LSTM for dual-avoided crossing model.

| Definition | Scaled |
|---|---|
| $x$ | $0.05x$ |
| $v$ | $(2000v-17)/30$ |
| $E_0$ | $20E_0$ |
| $E_1$ | $20E_1$ |
| $\rho_{00}$ | $\rho_{00}-0.5$ |
| $\text{Re}(\rho_{01})$ | $\text{Re}(\rho_{01})$ |
| $\text{Im}(\rho_{01})$ | $\text{Im}(\rho_{01})$ |

Table S5. Scaled input features of LSTM for extended coupling model.

| Definition | Scaled |
|---|---|
| $x$ | $0.05x$ |
| $v$ | $(2000v-17)/30$ |
| $E_0$ | $2.5E_0$ |
| $E_1$ | $2.5E_1$ |
| $\rho_{00}$ | $\rho_{00}-0.5$ |
| $\text{Re}(\rho_{01})$ | $\text{Re}(\rho_{01})$ |
| $\text{Im}(\rho_{01})$ | $\text{Im}(\rho_{01})$ |
| $\dfrac{G_0|v|}{(E_1-E_0)^2}$ | $\dfrac{G_0|v|}{3.75(E_1-E_0)^2}$ |
| $\dfrac{G_1|v|}{(E_1-E_0)^2}$ | $\dfrac{G_1|v|}{3.75(E_1-E_0)^2}$ |

# Section S2. Additional results of LSTM-FSSH

## S2.1 Performance of LSTM networks on testing sets

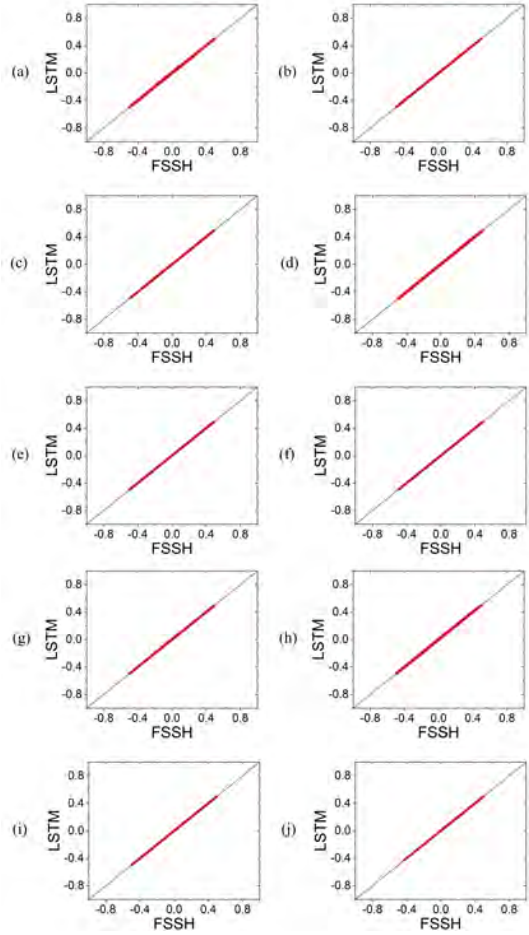

Figure S1: Comparison of LSTM predictions of electronic density matrix with reference for testing sets: single-avoided crossing model in the main text (a), dual-avoided crossing model in the main text (b), extended coupling model with multiple networks in the main text (c-f), dual-avoided crossing model with unsatisfactory performance in Fig. S17 (g), extended coupling model with unsatisfactory performance in Figs. S19 (h) and S14 (i), and extended coupling model with moderate performance in Fig. S12 (j). All values of $\rho_{00}$ are shifted by $-0.5$.

## S2.2 Time-dependent population for selected trajectories

We further analysis all trajectories that end in the same channel as the representative trajectories shown in Figs. 5, 6 and 7 in the main text. The number of trajectories between $-25.0 < x < 25.0$ in the lower surface is plotted as function of time in Figs. S2, S3 and S4 for the single-avoided crossing, dual-avoided crossing and extended coupling models, respectively. The results with LSTM-FSSH are consistent with reference. For example, the representative trajectory in Fig. 5(c) ends in T2 and corresponds to Fig. S2(c). The total number in T2 channel is 1350 using LSTM and 1450 in reference. All of these trajectories stay in the lower surface at the beginning. At step 1300, most trajectories switch to the upper surface, which take place a little latter using LSTM. For the dual-avoided crossing model, the representative trajectory in Fig. 6(c) ends in T1 and corresponds to Fig. S3(c). All of these trajectories evolve in the lower surface until $x > 25.0$ (see the sudden jump to zero in the end of simulation). An excited-state population lasting for a short time around step 800 can be captured with both methods. For the extended coupling model, the representative trajectory in Fig. 7(c) ends in T1 and corresponds to Fig. S4(c). It can be inferred that some trajectories at a high speed switch to the upper surface and hop back after step 2000 during LSTM-FSSH simulations, leading to a larger number of trajectories ending in T1 in comparison of the reference. This is consistent with the unphysical re-hops as discussed later.

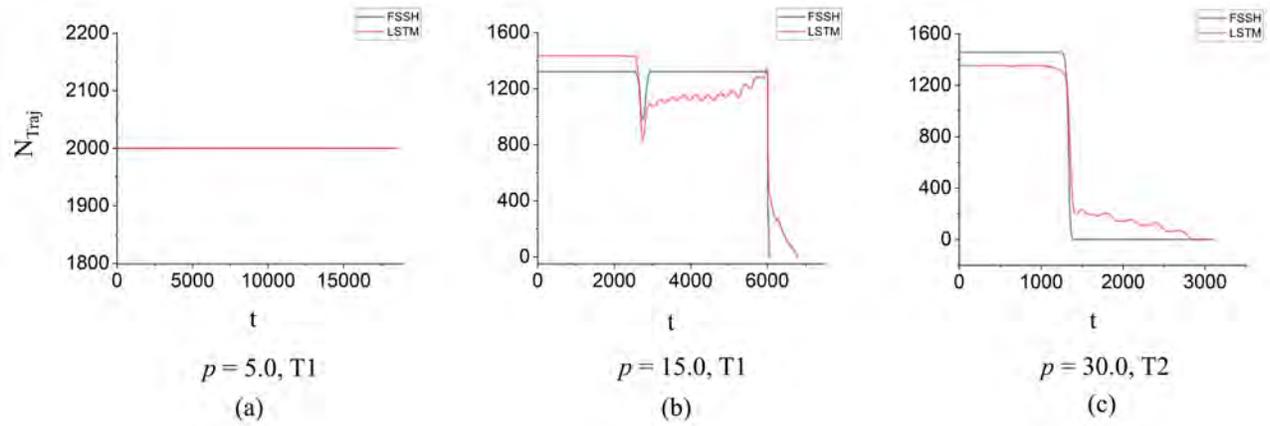

Figure S2: Number of trajectories between $-25.0 < x < 25.0$ in the lower surface plotted as function of time for single-avoided crossing model at different initial momenta and different channels: low speed/ending in T1 (a), medium speed/ending in T1 (b) and high speed/ending in T2 (c). Different colors represent different simulation methods (black: original FSSH as reference; red: LSTM-FSSH).

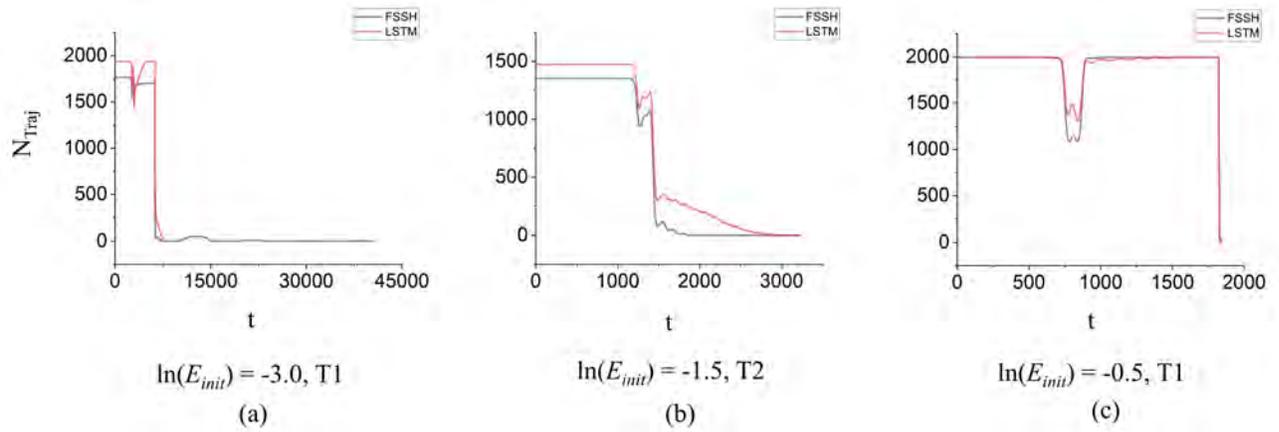

Figure S3: Number of trajectories between $-25.0 < x < 25.0$ in the lower surface plotted as function of time for dual-avoided crossing model at different initial momenta and different channels: low speed/ending in T1 (a), medium speed/ending in T2 (b) and high speed/ending in T1 (c). Different colors represent different simulation methods (black: original FSSH as reference; red: LSTM-FSSH).

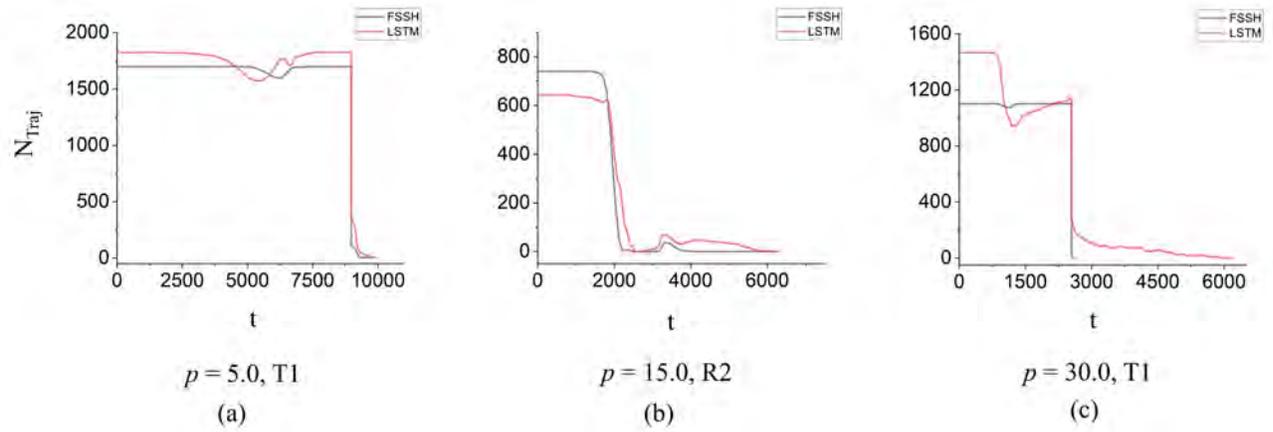

Figure S4: Number of trajectories between $-25.0 < x < 25.0$ in the lower surface plotted as function of time for extended coupling model at different initial momenta and different channels: low speed/ending in T1 (a), medium speed/ending in R2 (b) and high speed/ending in T1 (c). Different colors represent different simulation methods (black: original FSSH as reference; red: LSTM-FSSH).

## S2.3 Time and space distributions of hopping events

The time and space distributions of hopping events at a low, medium and high initial momentum are shown in Figs. S5, S7 and S9 for the single-avoided crossing model, dual-avoided crossing model and extended coupling model, respectively. We further distinguish the events between the first-hops and re-hops and display the results in Figs. S6, S8 and S10. In general, LSTM-FSSH is consistent with the reference qualitatively, but some discrepancies cannot be neglected. For the single-avoided crossing model, LSTM predicts more hopping events from the upper to lower surface (so-called down-hops). Most of them appear when the trajectory moves away from the coupling region, indicating an unphysical hopping. The LSTM-FSSH prediction on re-hops around the coupling region ($x = 0.0$) agrees with the reference, accompanied by unphysical re-hops. However, the contribution of unphysical hopping events has a small proportion ($< 10\%$) compared to all hops, so its impact on the final collective result seems little. The performance on the dual-avoided crossing model is better than others. For the extended coupling model, LSTM predicts some unphysical down-hops at a high speed. The relevant re-hops almost disappear using the original FSSH but can be produced with LSTM in the noncoupling region. As a result, the ratios of T1 channel at the same initial momentum ($p = 30.0$) are different using two methods. LSTM prediction at a low speed is not accurate for this model, which may originate from the disagreement of $\rho_{00}$ using two methods (see the representative trajectory in Fig. 7(a), step 6000-7000).

Two points should be considered to understand the performance of LSTM-FSSH. The first one is the cumulative error after a long-time dynamics simulation, which is unavoidable for any time-series ML predictor. However, the nuclei always evolve on a single potential energy surface during FSSH simulations, which reduces the impact of inaccurate electronic propagation on nuclear motion. In addition, FSSH requires a large number of trajectories to obtain collective results. It means that ML is useful to generate trajectory ensemble based on a small number of reference trajectories even if the length of trajectories is similar. Therefore, the cumulative error is not a serious issue at least in our present cases. The second one is

the re-hops that may be heavily influenced by cumulative errors. Surprisingly, we observe the agreement between the original FSSH and LSTM-FSSH when the trajectory crosses the coupling region again. LSTM actually predicts more re-hops far from the coupling region. It is possible to avoid such unphysical re-hops by rejecting surface hopping when the energy gap is larger than a certain threshold.

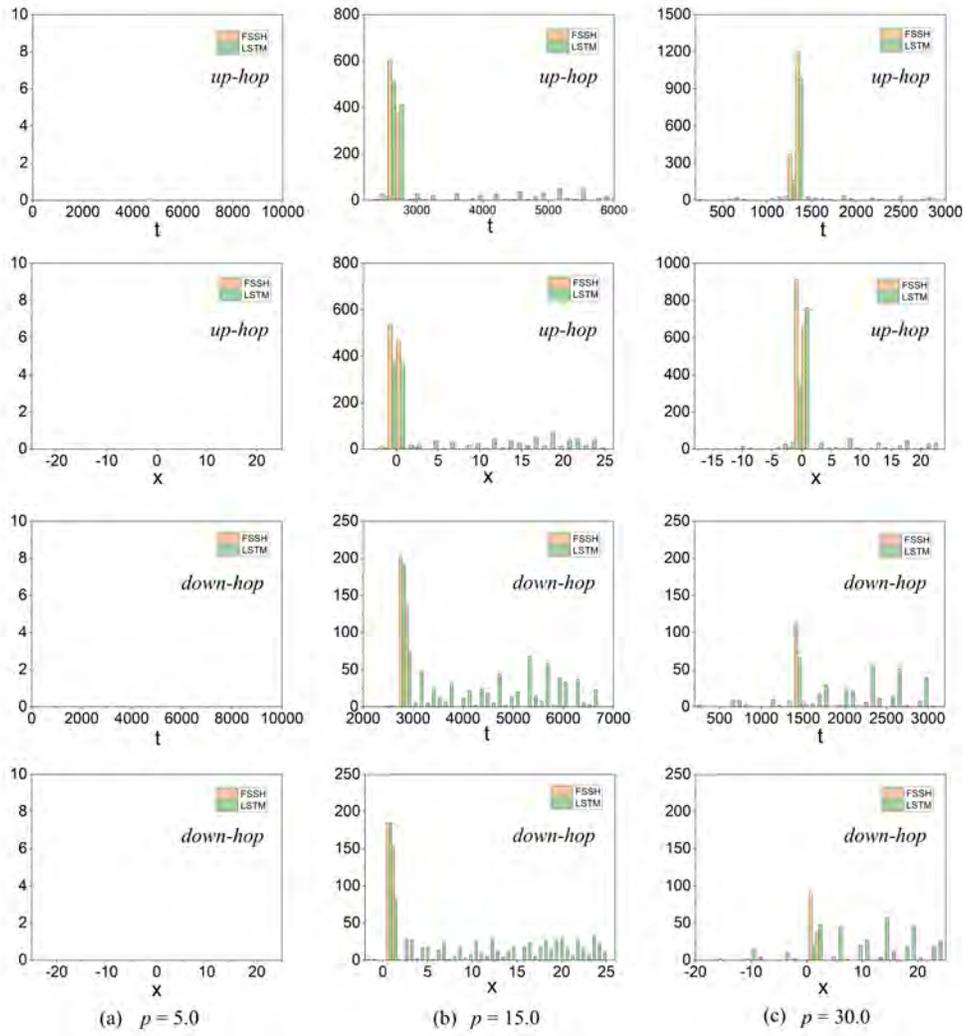

Figure S5: (from top to bottom) Time distribution of switching from lower to upper surface (i.e., up-hops), space distribution of up-hops, time distribution of switching from upper to lower surface (i.e., down-hops), and space distribution of down-hops in all trajectories for single-avoided crossing model at a low (a), medium (b) and high (c) initial momentum. Different colors represent different methods (orange: original FSSH as reference; green: LSTM-FSSH.)

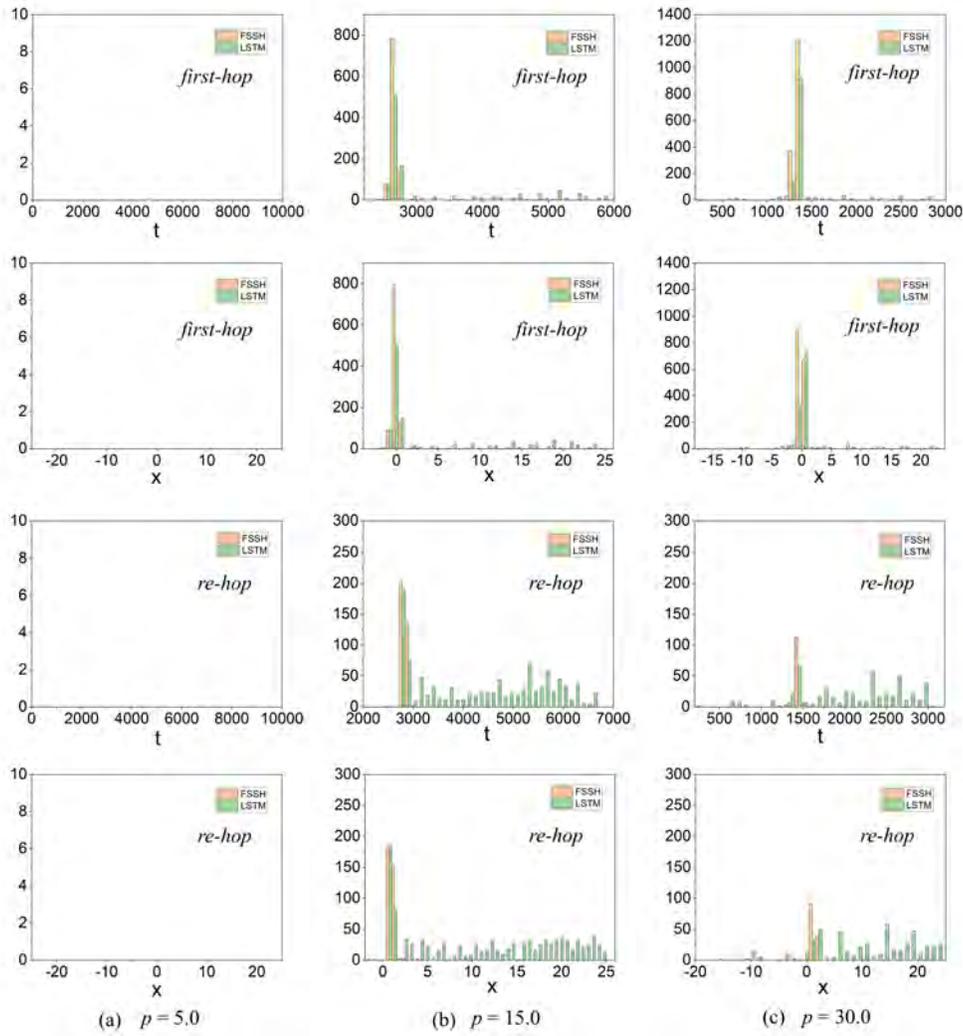

Figure S6: (from top to bottom) Time distribution of first-hops, space distribution of first-hops, time distribution of re-hops, and space distribution of re-hops taking place in all trajectories for single-avoided crossing model at a low (a), medium (b) and high (c) initial momentum. Different colors represent different methods (orange: original FSSH as reference; green: LSTM-FSSH.)

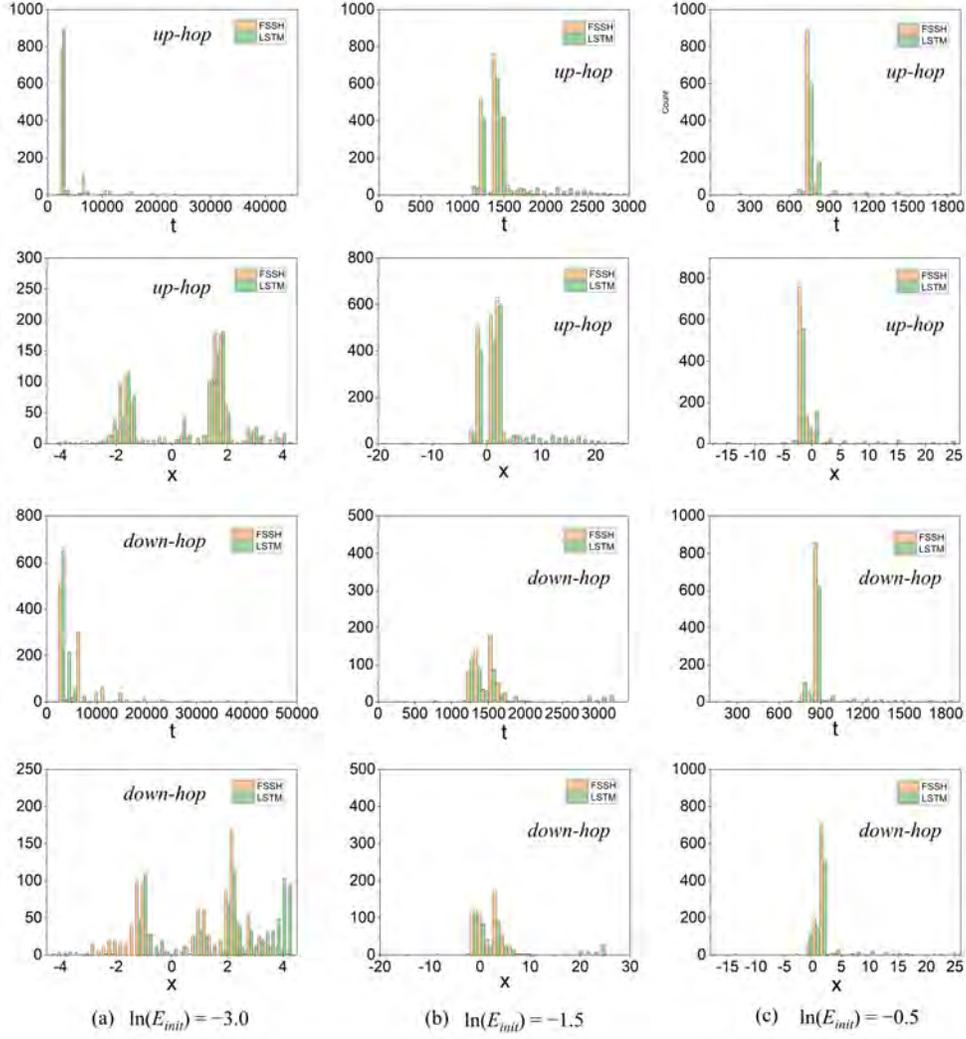

Figure S7: (from top to bottom) Time distribution of up-hops, space distribution of up-hops, time distribution of down-hops, and space distribution of down-hops in all trajectories for dual-avoided crossing model at a low (a), medium (b) and high (c) initial momentum. Different colors represent different methods (orange: original FSSH as reference; green: LSTM-FSSH.)

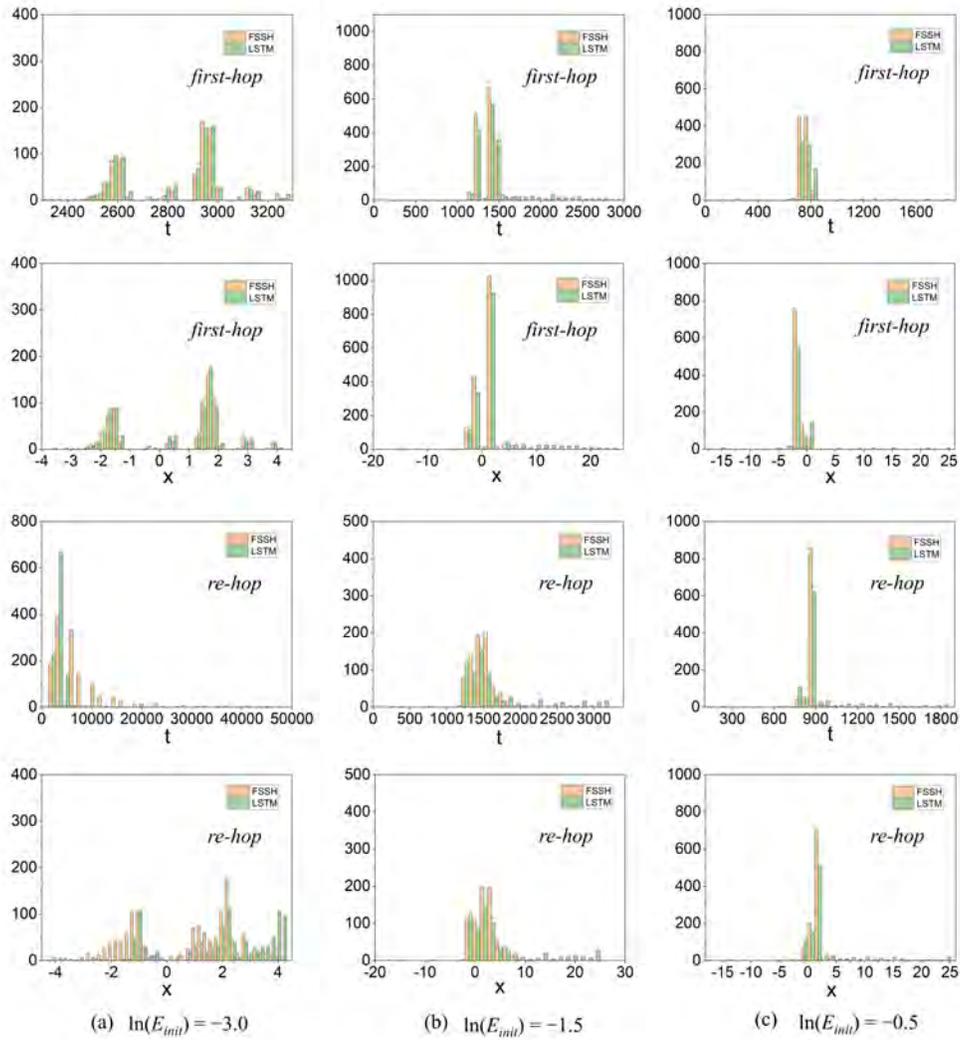

Figure S8: (from top to bottom) Time distribution of first-hops, space distribution of first-hops, time distribution of re-hops, and space distribution of re-hops taking place in all trajectories for dual-avoided crossing model at a low (a), medium (b) and high (c) initial momentum. Different colors represent different methods (orange: original FSSH as reference; green: LSTM-FSSH.)

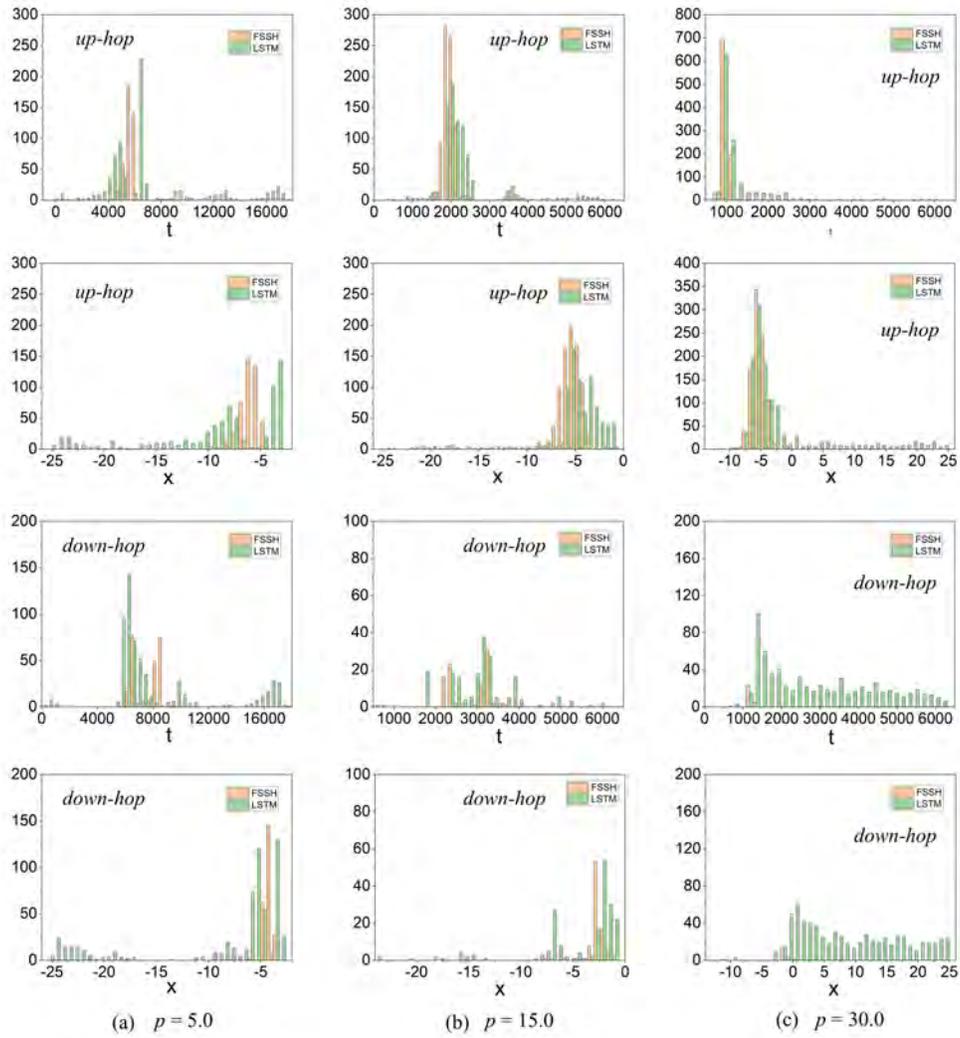

Figure S9: (from top to bottom) Time distribution of up-hops, space distribution of up-hops, time distribution of down-hops, and space distribution of down-hops in all trajectories for extended coupling model at a low (a), medium (b) and high (c) initial momentum. Different colors represent different methods (orange: original FSSH as reference; green: LSTM-FSSH.)

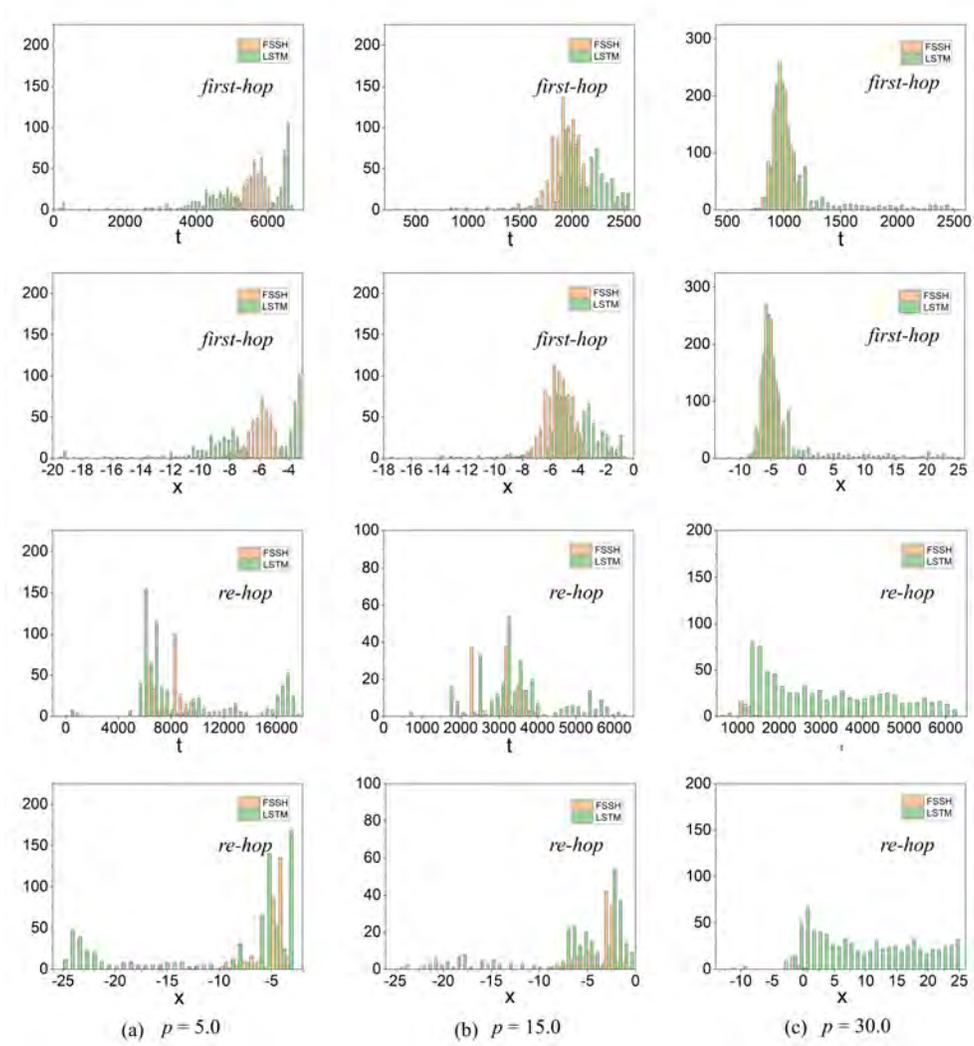

Figure S10: (from top to bottom) Time distribution of first-hops, space distribution of first-hops, time distribution of re-hops, and space distribution of re-hops taking place in all trajectories for extended coupling model at a low (a), medium (b) and high (c) initial momentum. Different colors represent different methods (orange: original FSSH as reference; green: LSTM-FSSH.)

# Section 3. More information on LSTM networks for extended coupling model

## S3.1 Multiple connected LSTM networks with trajectories in different cases

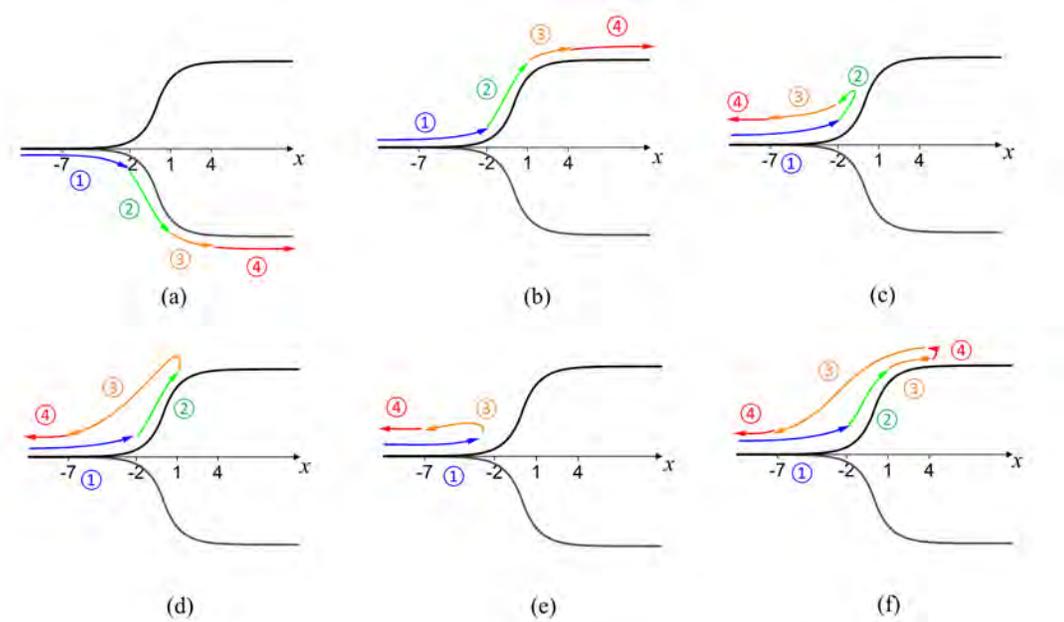

Figure S11: Multiple connected LSTM networks (labeled as 1 in blue, 2 in green, 3 in orange, and 4 in red) are applied and switched during propagation on extended coupling model. Different initial momenta or random surface hopping events lead to different cases represented in (a-f).

## S3.2 Attempts to construct a single LSTM network

We select the classical position $x$, velocity $v$, a function of nuclear degrees of freedom as $\frac{G_0|v|}{(E_1-E_0)^2}$ and $\frac{G_1|v|}{(E_1-E_0)^2}$, as well as $\rho_{00}$ as input features to construct a single LSTM network. Totally 165693 sequences are extracted from 50 reference trajectories with $M = 50$ and $N = 100$. The number of nodes for each LSTM layer and the dense layer are set as 48. The batch size for minimization is 64 with a learning rate as 0.001. As shown in Fig. S12, the time evolution of $\rho_{00}$ for three representative trajectories agree well with the reference. The collective results at low and high momenta match the original FSSH simulation, but the alternation of R1 and R2 populations at medium momenta disappears in LSTM-FSSH (see Fig. S13). Here the off-diagonal elements of electronic density matrix are ignored. Although the hopping probability can be obtained in absence of $\rho_{01}$ using Eq (15) in the main text, the phase of this term may strongly affect the population of reflection channels and should be included in ML models via $\operatorname{Re}(\rho_{01})$ and $\operatorname{Im}(\rho_{01})$. However, all attempts, in which we select $\frac{G_0|v|}{(E_1-E_0)^2}$, $\frac{G_1|v|}{(E_1-E_0)^2}$, $\rho_{00}$, $\operatorname{Re}(\rho_{01})$, $\operatorname{Im}(\rho_{01})$ and some other classical variables as the input features of a single LSTM network, fail in step (2d) (see an example in Figs. S14 and S15).

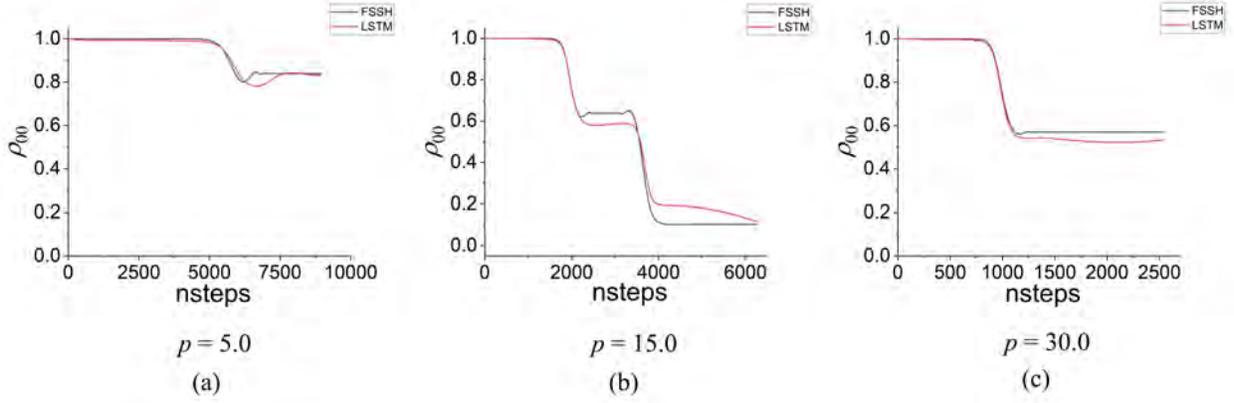

Figure S12: Electronic density matrix plotted as function of time for representative trajectories of extended coupling model at a low (a), medium (b) and high (c) initial momentum. LSTM network is constructed with input features as $x$, $v$, $\frac{G_0|v|}{(E_1-E_0)^2}$, $\frac{G_1|v|}{(E_1-E_0)^2}$ and $\rho_{00}$; all hyperparameters are listed in Section S3.2. Different colors represent different simulation methods (black: original FSSH as reference; red: LSTM-FSSH).

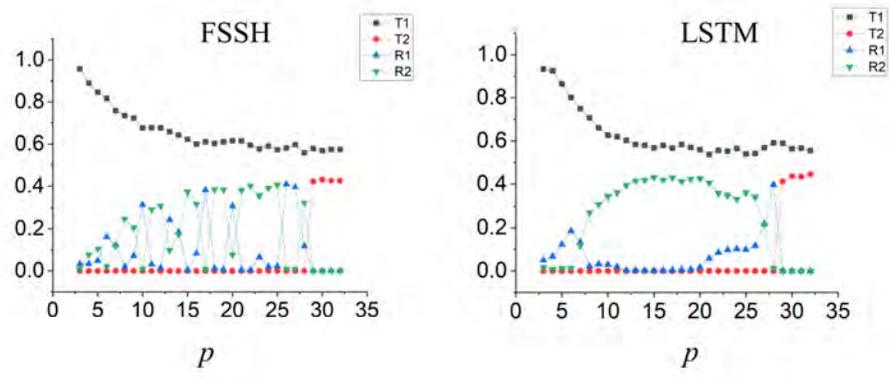

Figure S13: Collective result of extended coupling model obtained using LSTM in Fig. S12 in comparison with reference.

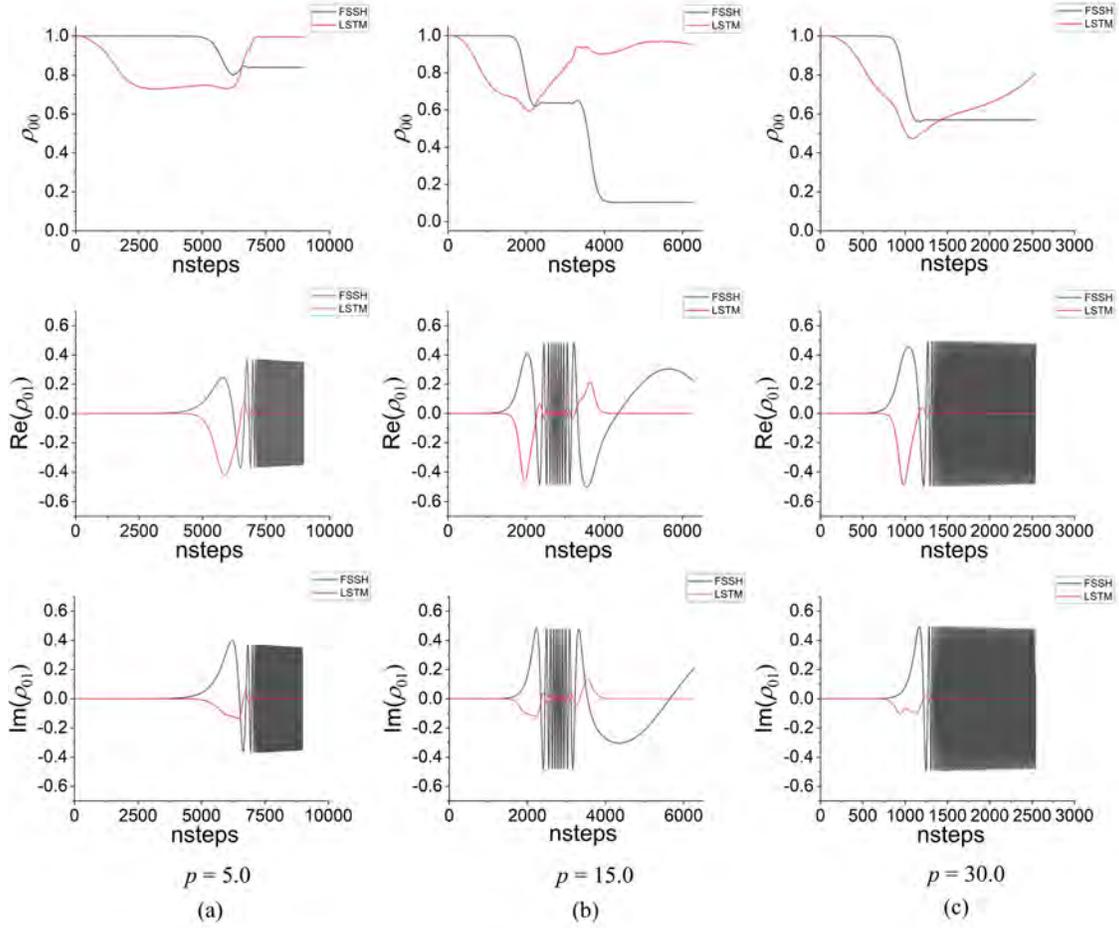

Figure S14: Electronic density matrix plotted as function of time for representative trajectories of extended coupling model at a low (a), medium (b) and high (c) initial momentum. LSTM is built with input features as $x$, $v$, $E_0$, $E_1$, $\frac{G_0|v|}{(E_1-E_0)^2}$, $\frac{G_1|v|}{(E_1-E_0)^2}$, $\rho_{00}$, $\text{Re}(\rho_{01})$ and $\text{Im}(\rho_{01})$; 417096 sequences in database are extracted with $M = 20$ and $N = 50$; the number of nodes for each layer is 72 with a batch size of 32 and a learning rate of 0.001. Different colors represent different simulation methods (black: original FSSH as reference; red: LSTM-FSSH).

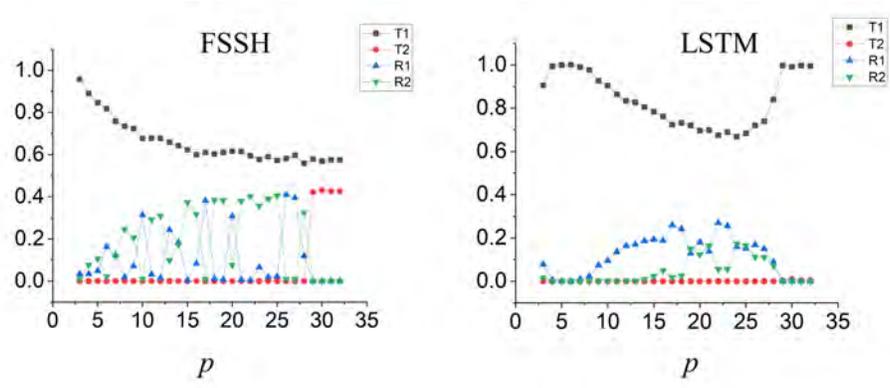

Figure S15: Collective result of extended coupling model obtained using LSTM in Fig. S14 in comparison with reference.

# S3.3 Performance of LSTM networks using more complex input features on other models

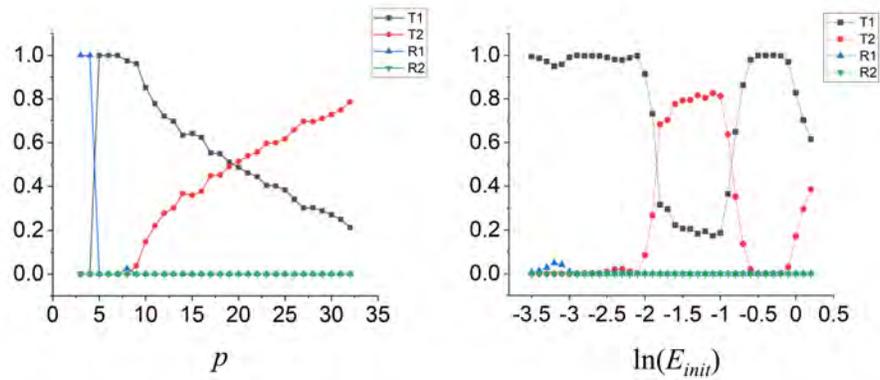

Figure S16: Collective results of LSTM-FSSH. The input features of LSTM for the extended coupling model in the main text are applied to single-avoided crossing (left) and dual-avoided crossing (right) models.

# Section 4. Other LSTM networks with unsatisfactory performance

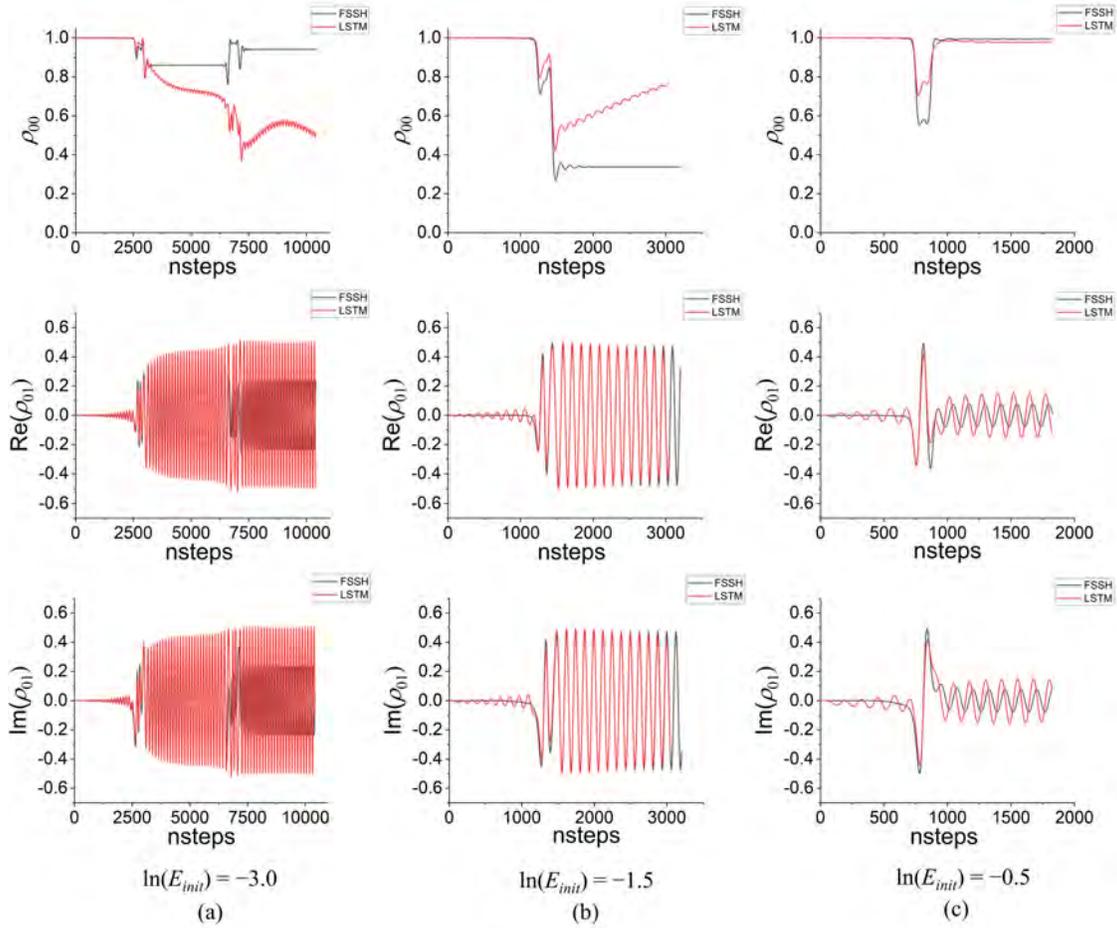

Figure S17: Electronic density matrix plotted as function of time for representative trajectories of dual-avoided crossing model at a low (a), medium (b) and high (c) initial momentum. LSTM is built with the number of nodes for each layer as 32; other hyperparameters are the same as that used in the main text. Different colors represent different simulation methods (black: original FSSH as reference; red: LSTM-FSSH).

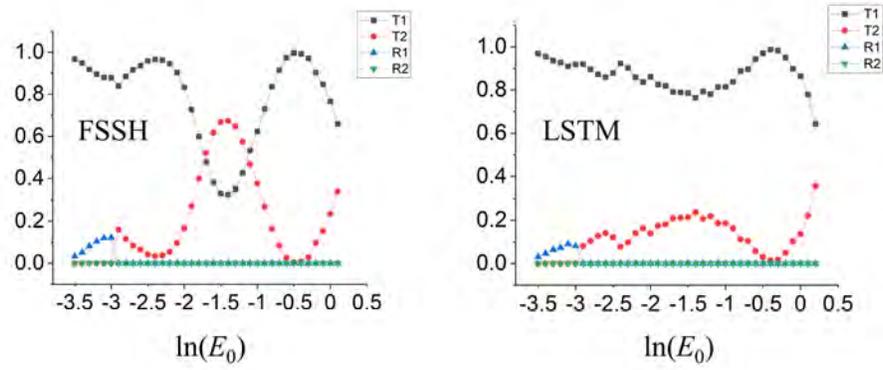

Figure S18: Collective result of dual-avoided crossing model obtained using LSTM in Fig. S17 in comparison with reference.

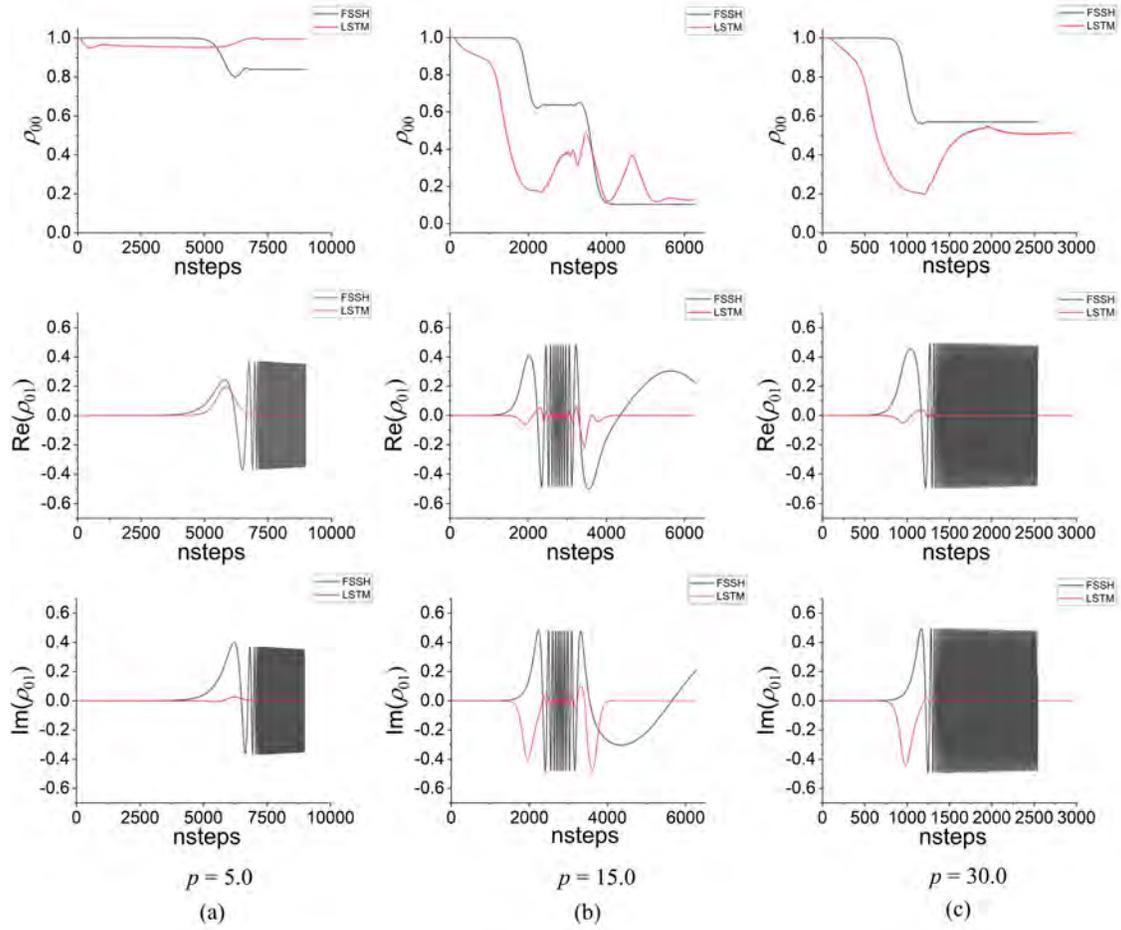

Figure S19: Electronic density matrix plotted as function of time for representative trajectories of extended coupling model at a low (a), medium (b) and high (c) initial momentum. LSTM is built with the same input features as the first and second test systems; 422341 sequences in database are extracted with $M = 20$ and $N = 50$; the number of nodes for each layer is 72 with a batch size of 64 and a learning rate of 0.001. Different colors represent different simulation methods (black: original FSSH as reference; red: LSTM-FSSH).

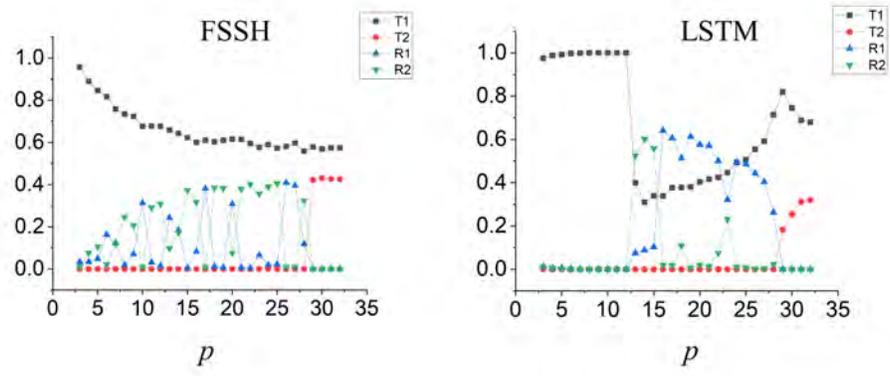

Figure S20: Collective result of extended coupling model obtained using LSTM in Fig. S19 in comparison with reference.

 

\end{document}